%
\def \preprint{Y}     
%
%
\if \preprint Y \input epsf \fi
%
\magnification=1200
\overfullrule=0pt
\baselineskip=15pt
\vsize=22truecm
\hsize=15truecm
\pageno=0

\font\titlefont=cmbx10 scaled \magstep1
\font\sectnfont=cmbx8  scaled \magstep1
\def\mname{\ifcase\month\or January \or February \or March \or April
           \or May \or June \or July \or August \or September
           \or October \or November \or December \fi}
\def\date{\hbox{\strut\mname \number\year}}
%
%
%
\newcount\FIGURENUMBER\FIGURENUMBER=0
\def\FIG#1{\expandafter\ifx\csname FG#1\endcsname\relax
               \global\advance\FIGURENUMBER by 1
               \expandafter\xdef\csname FG#1\endcsname
                              {\the\FIGURENUMBER}\fi}
\def\figtag#1{\expandafter\ifx\csname FG#1\endcsname\relax
               \global\advance\FIGURENUMBER by 1
               \expandafter\xdef\csname FG#1\endcsname
                              {\the\FIGURENUMBER}\fi
              \csname FG#1\endcsname\relax}
\def\fig#1{\expandafter\ifx\csname FG#1\endcsname\relax
               \global\advance\FIGURENUMBER by 1
               \expandafter\xdef\csname FG#1\endcsname
                      {\the\FIGURENUMBER}\fi
           Fig.~\csname FG#1\endcsname\relax}
\def\figand#1#2{\expandafter\ifx\csname FG#1\endcsname\relax
               \global\advance\FIGURENUMBER by 1
               \expandafter\xdef\csname FG#1\endcsname
                      {\the\FIGURENUMBER}\fi
           \expandafter\ifx\csname FG#2\endcsname\relax
               \global\advance\FIGURENUMBER by 1
               \expandafter\xdef\csname FG#2\endcsname
                      {\the\FIGURENUMBER}\fi
           figures \csname FG#1\endcsname\ and
                   \csname FG#2\endcsname\relax}
\def\figto#1#2{\expandafter\ifx\csname FG#1\endcsname\relax
               \global\advance\FIGURENUMBER by 1
               \expandafter\xdef\csname FG#1\endcsname
                      {\the\FIGURENUMBER}\fi
           \expandafter\ifx\csname FG#2\endcsname\relax
               \global\advance\FIGURENUMBER by 1
               \expandafter\xdef\csname FG#2\endcsname
                      {\the\FIGURENUMBER}\fi
           figures \csname FG#1\endcsname--\csname FG#2\endcsname\relax}
\newcount\TABLENUMBER\TABLENUMBER=0
\def\TABLE#1{\expandafter\ifx\csname TB#1\endcsname\relax
               \global\advance\TABLENUMBER by 1
               \expandafter\xdef\csname TB#1\endcsname
                          {\the\TABLENUMBER}\fi}
\def\tabletag#1{\expandafter\ifx\csname TB#1\endcsname\relax
               \global\advance\TABLENUMBER by 1
               \expandafter\xdef\csname TB#1\endcsname
                          {\the\TABLENUMBER}\fi
             \csname TB#1\endcsname\relax}
\def\table#1{\expandafter\ifx\csname TB#1\endcsname\relax
               \global\advance\TABLENUMBER by 1
               \expandafter\xdef\csname TB#1\endcsname{\the\TABLENUMBER}\fi
             Table \csname TB#1\endcsname\relax}
\def\tableand#1#2{\expandafter\ifx\csname TB#1\endcsname\relax
               \global\advance\TABLENUMBER by 1
               \expandafter\xdef\csname TB#1\endcsname{\the\TABLENUMBER}\fi
             \expandafter\ifx\csname TB#2\endcsname\relax
               \global\advance\TABLENUMBER by 1
               \expandafter\xdef\csname TB#2\endcsname{\the\TABLENUMBER}\fi
             Tables \csname TB#1\endcsname{} and
                    \csname TB#2\endcsname\relax}
\def\tableto#1#2{\expandafter\ifx\csname TB#1\endcsname\relax
               \global\advance\TABLENUMBER by 1
               \expandafter\xdef\csname TB#1\endcsname{\the\TABLENUMBER}\fi
             \expandafter\ifx\csname TB#2\endcsname\relax
               \global\advance\TABLENUMBER by 1
               \expandafter\xdef\csname TB#2\endcsname{\the\TABLENUMBER}\fi
            Tables \csname TB#1\endcsname--\csname TB#2\endcsname\relax}
\newcount\REFERENCENUMBER\REFERENCENUMBER=0
\def\REF#1{\expandafter\ifx\csname RF#1\endcsname\relax
               \global\advance\REFERENCENUMBER by 1
               \expandafter\xdef\csname RF#1\endcsname
                         {\the\REFERENCENUMBER}\fi}
\def\reftag#1{\expandafter\ifx\csname RF#1\endcsname\relax
               \global\advance\REFERENCENUMBER by 1
               \expandafter\xdef\csname RF#1\endcsname
                      {\the\REFERENCENUMBER}\fi
             \csname RF#1\endcsname\relax}
\def\ref#1{\expandafter\ifx\csname RF#1\endcsname\relax
               \global\advance\REFERENCENUMBER by 1
               \expandafter\xdef\csname RF#1\endcsname
                      {\the\REFERENCENUMBER}\fi
             [\csname RF#1\endcsname]\relax}
\def\refto#1#2{\expandafter\ifx\csname RF#1\endcsname\relax
               \global\advance\REFERENCENUMBER by 1
               \expandafter\xdef\csname RF#1\endcsname
                      {\the\REFERENCENUMBER}\fi
           \expandafter\ifx\csname RF#2\endcsname\relax
               \global\advance\REFERENCENUMBER by 1
               \expandafter\xdef\csname RF#2\endcsname
                      {\the\REFERENCENUMBER}\fi
             [\csname RF#1\endcsname--\csname RF#2\endcsname]\relax}
\def\refand#1#2{\expandafter\ifx\csname RF#1\endcsname\relax
               \global\advance\REFERENCENUMBER by 1
               \expandafter\xdef\csname RF#1\endcsname
                      {\the\REFERENCENUMBER}\fi
                \expandafter\ifx\csname RF#2\endcsname\relax
               \global\advance\REFERENCENUMBER by 1
               \expandafter\xdef\csname RF#2\endcsname
                      {\the\REFERENCENUMBER}\fi
            [\csname RF#1\endcsname{} and \csname RF#2\endcsname]\relax}
\def\refs#1#2{\expandafter\ifx\csname RF#1\endcsname\relax
               \global\advance\REFERENCENUMBER by 1
               \expandafter\xdef\csname RF#1\endcsname
                      {\the\REFERENCENUMBER}\fi
           \expandafter\ifx\csname RF#2\endcsname\relax
               \global\advance\REFERENCENUMBER by 1
               \expandafter\xdef\csname RF#2\endcsname
                      {\the\REFERENCENUMBER}\fi
            [\csname RF#1\endcsname,\csname RF#2\endcsname]\relax}
\def\refss#1#2#3{\expandafter\ifx\csname RF#1\endcsname\relax
               \global\advance\REFERENCENUMBER by 1
               \expandafter\xdef\csname RF#1\endcsname
                      {\the\REFERENCENUMBER}\fi
           \expandafter\ifx\csname RF#2\endcsname\relax
               \global\advance\REFERENCENUMBER by 1
               \expandafter\xdef\csname RF#2\endcsname
                      {\the\REFERENCENUMBER}\fi
           \expandafter\ifx\csname RF#3\endcsname\relax
               \global\advance\REFERENCENUMBER by 1
               \expandafter\xdef\csname RF#3\endcsname
                      {\the\REFERENCENUMBER}\fi
[\csname RF#1\endcsname,\csname RF#2\endcsname,\csname RF#3\endcsname]\relax}
\def\refsss#1#2#3#4{\expandafter\ifx\csname RF#1\endcsname\relax
               \global\advance\REFERENCENUMBER by 1
               \expandafter\xdef\csname RF#1\endcsname
                      {\the\REFERENCENUMBER}\fi
           \expandafter\ifx\csname RF#2\endcsname\relax
               \global\advance\REFERENCENUMBER by 1
               \expandafter\xdef\csname RF#2\endcsname
                      {\the\REFERENCENUMBER}\fi
           \expandafter\ifx\csname RF#3\endcsname\relax
               \global\advance\REFERENCENUMBER by 1
               \expandafter\xdef\csname RF#3\endcsname
                      {\the\REFERENCENUMBER}\fi
           \expandafter\ifx\csname RF#4\endcsname\relax
               \global\advance\REFERENCENUMBER by 1
               \expandafter\xdef\csname RF#4\endcsname
                      {\the\REFERENCENUMBER}\fi
[\csname RF#1\endcsname,\csname RF#2\endcsname,\csname RF#3\endcsname,\csname
RF#4\endcsname]\relax}
\def\refssss#1#2#3#4#5{\expandafter\ifx\csname RF#1\endcsname\relax
               \global\advance\REFERENCENUMBER by 1
               \expandafter\xdef\csname RF#1\endcsname
                      {\the\REFERENCENUMBER}\fi
           \expandafter\ifx\csname RF#2\endcsname\relax
               \global\advance\REFERENCENUMBER by 1
               \expandafter\xdef\csname RF#2\endcsname
                      {\the\REFERENCENUMBER}\fi
           \expandafter\ifx\csname RF#3\endcsname\relax
               \global\advance\REFERENCENUMBER by 1
               \expandafter\xdef\csname RF#3\endcsname
                      {\the\REFERENCENUMBER}\fi
           \expandafter\ifx\csname RF#4\endcsname\relax
               \global\advance\REFERENCENUMBER by 1
               \expandafter\xdef\csname RF#4\endcsname
                      {\the\REFERENCENUMBER}\fi
           \expandafter\ifx\csname RF#5\endcsname\relax
               \global\advance\REFERENCENUMBER by 1
               \expandafter\xdef\csname RF#5\endcsname
                      {\the\REFERENCENUMBER}\fi
[\csname RF#1\endcsname,\csname RF#2\endcsname,\csname RF#3\endcsname,\csname
RF#4\endcsname,\csname RF#5\endcsname]\relax}
\def\Ref#1{\expandafter\ifx\csname RF#1\endcsname\relax
               \global\advance\REFERENCENUMBER by 1
               \expandafter\xdef\csname RF#1\endcsname
                      {\the\REFERENCENUMBER}\fi
             Ref.~\csname RF#1\endcsname\relax}
\def\Refs#1#2{\expandafter\ifx\csname RF#1\endcsname\relax
               \global\advance\REFERENCENUMBER by 1
               \expandafter\xdef\csname RF#1\endcsname
                      {\the\REFERENCENUMBER}\fi
           \expandafter\ifx\csname RF#2\endcsname\relax
               \global\advance\REFERENCENUMBER by 1
               \expandafter\xdef\csname RF#2\endcsname
                      {\the\REFERENCENUMBER}\fi
        Refs.~\csname RF#1\endcsname{},\csname RF#2\endcsname\relax}
\def\Refto#1#2{\expandafter\ifx\csname RF#1\endcsname\relax
               \global\advance\REFERENCENUMBER by 1
               \expandafter\xdef\csname RF#1\endcsname
                      {\the\REFERENCENUMBER}\fi
           \expandafter\ifx\csname RF#2\endcsname\relax
               \global\advance\REFERENCENUMBER by 1
               \expandafter\xdef\csname RF#2\endcsname
                      {\the\REFERENCENUMBER}\fi
            Refs.~\csname RF#1\endcsname--\csname RF#2\endcsname]\relax}
\def\Refand#1#2{\expandafter\ifx\csname RF#1\endcsname\relax
               \global\advance\REFERENCENUMBER by 1
               \expandafter\xdef\csname RF#1\endcsname
                      {\the\REFERENCENUMBER}\fi
           \expandafter\ifx\csname RF#2\endcsname\relax
               \global\advance\REFERENCENUMBER by 1
               \expandafter\xdef\csname RF#2\endcsname
                      {\the\REFERENCENUMBER}\fi
        Refs.~\csname RF#1\endcsname{} and \csname RF#2\endcsname\relax}
\newcount\EQUATIONNUMBER\EQUATIONNUMBER=0
\def\EQ#1{\expandafter\ifx\csname EQ#1\endcsname\relax
               \global\advance\EQUATIONNUMBER by 1
               \expandafter\xdef\csname EQ#1\endcsname
                          {\the\EQUATIONNUMBER}\fi}
\def\eqtag#1{\expandafter\ifx\csname EQ#1\endcsname\relax
               \global\advance\EQUATIONNUMBER by 1
               \expandafter\xdef\csname EQ#1\endcsname
                      {\the\EQUATIONNUMBER}\fi
            \csname EQ#1\endcsname\relax}
\def\EQNO#1{\expandafter\ifx\csname EQ#1\endcsname\relax
               \global\advance\EQUATIONNUMBER by 1
               \expandafter\xdef\csname EQ#1\endcsname
                      {\the\EQUATIONNUMBER}\fi
            \eqno(\csname EQ#1\endcsname)\relax}
\def\EQNM#1{\expandafter\ifx\csname EQ#1\endcsname\relax
               \global\advance\EQUATIONNUMBER by 1
               \expandafter\xdef\csname EQ#1\endcsname
                      {\the\EQUATIONNUMBER}\fi
            (\csname EQ#1\endcsname)\relax}
\def\eq#1{\expandafter\ifx\csname EQ#1\endcsname\relax
               \global\advance\EQUATIONNUMBER by 1
               \expandafter\xdef\csname EQ#1\endcsname
                      {\the\EQUATIONNUMBER}\fi
          Eq.~(\csname EQ#1\endcsname)\relax}
\def\eqand#1#2{\expandafter\ifx\csname EQ#1\endcsname\relax
               \global\advance\EQUATIONNUMBER by 1
               \expandafter\xdef\csname EQ#1\endcsname
                        {\the\EQUATIONNUMBER}\fi
          \expandafter\ifx\csname EQ#2\endcsname\relax
               \global\advance\EQUATIONNUMBER by 1
               \expandafter\xdef\csname EQ#2\endcsname
                      {\the\EQUATIONNUMBER}\fi
         Eqs.~(\csname EQ#1\endcsname) and (\csname EQ#2\endcsname)\relax}
\def\eqto#1#2{\expandafter\ifx\csname EQ#1\endcsname\relax
               \global\advance\EQUATIONNUMBER by 1
               \expandafter\xdef\csname EQ#1\endcsname
                      {\the\EQUATIONNUMBER}\fi
          \expandafter\ifx\csname EQ#2\endcsname\relax
               \global\advance\EQUATIONNUMBER by 1
               \expandafter\xdef\csname EQ#2\endcsname
                      {\the\EQUATIONNUMBER}\fi
          Eqs.~\csname EQ#1\endcsname--\csname EQ#2\endcsname\relax}
%
\newcount\SECTIONNUMBER\SECTIONNUMBER=0
\newcount\SUBSECTIONNUMBER\SUBSECTIONNUMBER=0
\def\section#1{\global\advance\SECTIONNUMBER by 1\SUBSECTIONNUMBER=0
      \bigskip\goodbreak\line{{\sectnfont \the\SECTIONNUMBER.\ #1}\hfil}
      \smallskip}
\def\subsection#1{\global\advance\SUBSECTIONNUMBER by 1
      \bigskip\goodbreak\line{{\sectnfont
         \the\SECTIONNUMBER.\the\SUBSECTIONNUMBER.\ #1}\hfil}
      \smallskip}
%

\def\NP{{\sl Nucl.\ Phys.\ }}
\def\PL{{\sl Phys.\ Lett.\ }}
\def\PR{{\sl Phys.\ Rev.\ }}

\def\PRL{{\sl Phys.\ Rev.\ Lett.\ }}

\def\ZP{{\sl Z.\ Phys.\ }}
%
\def\lsim{\raise0.3ex\hbox{$<$\kern-0.75em\raise-1.1ex\hbox{$\sim$}}}
\def\gsim{\raise0.3ex\hbox{$>$\kern-0.75em\raise-1.1ex\hbox{$\sim$}}}
\def\caption#1{\setbox0=\hbox{\noindent\it Figure #1}
   	\setbox1=\vbox{\hsize=0.9\hsize\normalbaselines%
		\noindent\it Figure #1}
   \ifdim\wd0<0.9\hsize%
	\medskip
        \centerline{\box0}%
   \else%
	\medskip
        \centerline{\box1}%
   \fi%
}
%
\def\lsim{\raise0.3ex\hbox{$<$\kern-0.75em\raise-1.1ex\hbox{$\sim$}}}
\def\gsim{\raise0.3ex\hbox{$>$\kern-0.75em\raise-1.1ex\hbox{$\sim$}}}
%
\def\ie{\it{i.e.}\rm,\ }
\def\bar#1{\overline#1}
\def\GR{g_4}\def\gr{\ifmmode\GR\else$\GR~$\fi}
\def\BR{\beta_r}\def\br{\ifmmode\BR\else$\BR~$\fi}
\def\RAT{T_c/\sqrt{\sigma}}\def\rat{\ifmmode\RAT\else$\RAT~$\fi}
\def\LMS{\Lambda_{\rm{\bar{M}\bar{S}}}}
\def\lms{\ifmmode\LMS\else$\LMS~$\fi}
\def\TC{T_c/\lms}\def\tc{\ifmmode\TC\else$\TC~$\fi}
\def\STR{\sqrt{\sigma}/\lms}\def\str{\ifmmode\STR\else$\STR~$\fi}
\def\NT{N_{\tau}}\def\nt{\ifmmode\NT\else$\NT~$\fi}
\def\NTA{N_{\tau}=8}\def\nta{\ifmmode\NTA\else$\NTA~$\fi}
\def\NTB{N_{\tau}=16}\def\ntb{\ifmmode\NTB\else$\NTB~$\fi}
\def\MEV{{\rm MeV}}\def\mev{\ifmmode\MEV\else$\MEV~$\fi}
\def\NS{N_{\sigma}}\def\ns{\ifmmode\NS\else$\NS~$\fi}
\def\LAT{N_{\tau} \times N_{\sigma}^3}\def\lat{\ifmmode\LAT\else$\LAT~$\fi}

\def\NP{{\sl Nucl.\ Phys.\ }}
\def\PL{{\sl Phys.\ Lett.\ }}
\def\PR{{\sl Phys.\ Rev.\ }}

\def\PRL{{\sl Phys.\ Rev.\ Lett.\ }}

\def\ZP{{\sl Z.\ Phys.\ }}

\def\binum{\hbox{BI-TP 92-26 \strut}}
\def\scrinum{\hbox{FSU-SCRI-92-103 \strut}}
\def\hlrznum{\hbox{HLRZ-92-39 \strut}}

\def\banner{\hfill\hbox{\vbox{\offinterlineskip
                              \binum\hlrznum\scrinum}}\relax}
\def\manner{\hbox{\vbox{\offinterlineskip
                \binum\hlrznum\scrinum\hbox{\mname \number\year\ REVISED
\strut}
}}\hfill\relax}
\footline={\ifnum\pageno=0\manner\else\hfil\number\pageno\hfil\fi}
{\vsize=18truecm\banner\bigskip\baselineskip=15pt
\bigskip\bigskip
\begingroup\titlefont\obeylines
\hfil Scaling and Asymptotic Scaling in the SU(2) \hfil
\hfil Gauge Theory\hfil
\endgroup\bigskip
\centerline{
  J.~Fingberg{$^1$},
  U.~Heller{$^2$}
  and F.~Karsch{$^{1,3}$} }
\bigskip
\centerline{
$^1$~HLRZ, c/o KFA J\"ulich, D-5170 J\"ulich, Germany}
\centerline{
$^2$~SCRI, The Florida State University, Tallahassee, USA}
\centerline{
$^{~~3}$~Fakult\"at f\"ur Physik, Universit\"at Bielefeld,
        D-4800 Bielefeld 1, Germany}
\bigskip
\bigskip
\bigskip\bigskip\bigskip\centerline{\bf ABSTRACT}\medskip
We determine the critical couplings for the deconfinement phase transition
in $SU(2)$ gauge theory on \lat lattices with \nta and 16 and \ns varying
between 16 and 48. A comparison with string tension data shows scaling of
the ratio \rat in the entire coupling regime $\beta =2.30-2.75$, while the
individual quantities still exhibit large scaling violations. We find
\rat=0.69(2). We also discuss in detail the extrapolation of \tc and \str
to the continuum limit. Our result, which is consistent with the above
ratio, is $\tc = 1.23(11)$ and $\str = 1.79(12)$. We also comment upon
corresponding results for $SU(3)$ gauge theory and four flavour QCD.
\vfil\eject}

\noindent
{\bf 1. Introduction}

Ever since the pioneering work of M. Creutz \ref{creutz} the approach to
asymptotic scaling, and thus the continuum limit, was one of  the central
issues in studies of gauge theories on the lattice. Although the first
results were promising, it soon became clear that simulations on large
lattices are needed in order to establish asymptotic scaling for
asymptotically free quantum field theories such as QCD. In fact, recent
numerical studies of the $O(3)$ $\sigma$-model in two dimensions
\ref{wolff} suggest that the asymptotic scaling regime may not be reached
even for quite large correlation length while at the same time ratios of
physical observables show scaling behaviour.

The lack of asymptotic scaling as well as the scaling of certain ratios of
physical observables has also been observed in $SU(N)$ gauge theories. In
particular in the case of the $SU(3)$ gauge theory the deviations from
asymptotic scaling are large and have been noticed early as a dip in the
discrete $\beta$-function \refs{mcrga}{mcrgb}, which led to deviations from
asymptotic scaling by more than 50\% for certain values of the gauge
couplings. Although the dip is not that pronounced for $SU(2)$ gauge
theories, there are clear deviations from asymptotic scaling seen even for
the largest values of the coupling, $\beta=2N/g^2$ for colour group
$SU(N)$, studied so far. In particular the analysis of the heavy quark
potential and the string tension has been performed up to couplings as
large as $\beta=2.85$ \ref{pota}. Still there is no hint for asymptotic
scaling at this large $\beta$-value, which already corresponds to lattice
spacings as small as $\sim 0.05$~fm. Furthermore, an analysis of the short
distance part of the heavy quark potential suggests that the approach to
the continuum limit may even be as slow as in the two-dimensional
$\sigma$-model \ref{potb}.

One of the best studied quantities in $SU(2)$ gauge theory is the finite
temperature deconfinement phase transition. For lattices of size \lat with
$\nt \le 6$ the critical coupling has been determined with high accuracy
\refs{enga}{engb} and an extrapolation to spatially infinite volume could
be performed. Moreover, an analysis of the critical exponents at the
transition point were in perfect agreement with those of the
three-dimensional Ising model. Here it turned out that the Binder cumulant
of the order parameter is an observable which is well suited to locate the
critical coupling for given $N_{\tau}$ as finite spatial size corrections
are only due to the presence of irrelevant operators.

So far the analysis of the scaling of the ratio $T_c/\sqrt{\sigma}$ was
limited to a rather small coupling regime in the case of $SU(2)$ as the
critical couplings for the deconfinement transition have been determined
only for $\nt \le 6$. It is the purpose of this paper to further
investigate the scaling properties of the $SU(2)$ gauge theory. We will
extend earlier studies of the deconfinement transition to lattices up to a
temporal size $N_\tau=16$. This will enable us to perform a quantitative
test of scaling in $SU(2)$. Furthermore, we can follow the apparent scaling
violations over a large range of couplings, which allows us to analyze
various extrapolation schemes to extract \tc in the continuum limit.

This paper is organized as follows: In the next section we discuss our
strategy of calculating the critical couplings for the deconfinement
transition on lattices with large temporal extent. In particular we discuss
the finite-size scaling of cumulants of the Polyakov loop expectation
value. In section 3 we present our numerical results. Section 4 is devoted
to a discussion of scaling and asymptotic scaling and a detailed discussion
of the extrapolation of these results to the continuum limit. Finally
section 5 contains our conclusions.

\bigskip
\noindent
{\bf 2. Finite-Size Scaling and the Continuum Limit}

Usually finite-size scaling (FSS) in the vicinity of a finite temperature
phase transition is discussed for lattice $SU(N)$ gauge models, without
trying to make contact with the continuum limit, \ie the scaling
properties are studied on lattices of size $N_\tau \times N_\sigma^d$ with
fixed \nt and varying $N_\sigma$, where $d$ denotes the spatial dimension
and the model is viewed as a d-dimensional spin system.
In the continuum limit
the FSS properties of these non-abelian models should, of course, be
discussed in terms of the  physical volume, $V=L^d$, and the temperature,
$T$, in the vicinity of the deconfinement transition temperature $T_c$. We
will study here how the scaling behaviour of the continuum theory emerges
from the lattice free energy on arbitrary lattices, \ie when varying \nt
and $N_\sigma$.

For a continuum theory having a simple critical point
and a characteristic length $L=V^{1/d}$ the singular part of
the free energy density,
$$
f_s = {F_s \over {TV}} = - {{\rm{ln}~Z_s} \over {TV}} ~~,~~
\EQNO{fs_cont}
$$
\noindent
is described by a universal finite-size scaling
form \refs{barbera}{privmana},
$$
f_s(T,H;L) = L^{-d}Q_{f_s}\left(g_T L^{1\over \nu},g_H L^{{\beta +
\gamma}\over {\nu}}\right) ~~.~~
\EQNO{fs_fss}
$$
\noindent
Here we assume that corrections to scaling from irrelevant scaling fields
$g_i L^{y_i}$, proportional to negative powers of $L$, can be neglected;
for $N_\tau=6$ a value of $y_1=-0.9$ has been found for
the $SU(2)$ gauge theory \ref{engb},
showing that irrelevant contributions disappear rather fast with
increasing $N_\sigma$.

On a lattice of size $N_\tau \times N_\sigma^d$
the length scale $L$ and the temperature $T$ are
given in units of the lattice spacing
$$ \eqalignno{
L      &= N_\sigma a ~~,~~ & \EQNM{L} \cr
T^{-1} &= N_\tau a   ~~.~~ & \EQNM{T} \cr
             }
$$
In general the lattice spacing $a$ is a complicated function
of the coupling $\beta=2N/g^2$.
The dependence of the lattice spacing on $\beta$
is known only in the continuum limit
in the form of the renormalization group equation
$$
a \Lambda_L = \left({\beta \over {2Nb_0}}\right)^{b_1/2b_0^2}
                     \rm{exp}\left(- {\beta \over {4Nb_0}}\right) ~~.~~
\EQNO{alam}
$$
\noindent
Therefore it is advantageous to replace the length scale $L$ by
the dimensionless combination,
$$
L \cdot T = {N_\sigma \over N_\tau} ~~.~~ \EQNO{ratio}
$$
Using this ratio in the FSS relation for the singular part of the free
energy density we get
$$
f_s(t,h;{N_\sigma};{N_\tau}) = {\left(N_\sigma \over N_\tau\right)}^{-d}
Q_{f_s}\left(g_t \left({N_\sigma\over N_\tau}\right)^{1\over \nu},
g_h \left({N_\sigma\over N_\tau}\right)^{{\beta + \gamma}\over {\nu}}\right)
 ~~.~~ \EQNO{free}
$$
\noindent
The scaling function $Q_{f_s}$ depends on the temperature $T$ and the
external field strength $h$ through thermal and  magnetic
scaling fields,
$$\eqalignno{
g_t &= c_t t (1+b_t t) + O(th,t^3)        ~~,~~&\rm{\EQNM{gt}} \cr
g_h &= c_h h (1+b_h t) + O(th^2,t^2h,h^2) ~~,~~&\rm{\EQNM{gh}} \cr
            }
$$
\noindent
with non-universal metric coefficients $c_t$, $c_h$, $b_t$ and $b_h$ still
carrying a possible $N_\tau$ dependence. Here $t$ is the reduced
temperature, $t=(T-T_c)/T_c$, which in the neighbourhood of the transition
point can be approximated by
$$
t=\left(\beta-\beta_{c,\infty}\right) {1\over 4Nb_0}
\left[1 - {{2Nb_1}\over{b_0}} \beta_{c,\infty}^{-1} \right] ~~.~~
\EQNO{t}
$$
\noindent
This approximation reproduces the correct reduced  temperature in the
continuum limit, which is easily verified by using \eq{alam}. We note that
$t$ has $O(\beta_{c,\infty}^{-1})$ corrections to the leading term, which,
however, contribute less than 8\% in the relevant coupling regime,
\ie $\beta > 2.0$ for $SU(2)$ and $\beta > 5.5$ for $SU(3)$. This non-leading
term  introduces a logarithmic dependence of $t$, and thus $Q_{f_s}$, on
$N_\tau$ through \eqand{T}{alam}. A priori we cannot exclude that this
violation of the otherwise universal $N_\sigma/N_\tau$ dependence is
enhanced in the non-asymptotic scaling regime. However, as we will show
later we do not find any hints for this in our Monte Carlo data.

A non-vanishing magnetic field strength $h$ corresponds
to adding a symmetry breaking
term of the form $h Z(a,N_\tau) \ns^d P$ to the action.
Here $P$ denotes the Polyakov loop, defined as
$$
P=\ns ^{-d} \sum_{\vec x} \prod_{x_0 = 1}^{\nt} U_{(x_0, \vec x ),0}
 ~~,~~ \EQNO{pol}
$$
and $Z(a,N_\tau)$ is a renormalization factor necessary
to remove divergent self-energy contributions to the Polyakov loop,
which represents a static heavy quark source. A physical order
parameter, $<P_p>$, not vanishing in the continuum limit,
a susceptibility $\chi_p$ and a normalized fourth cumulant $g_4$
may then be defined through derivatives of $f_s$ with respect
to the external magnetic field strength $h$ at $h=0$,
$$
<P_p>={N_\tau}^d Z(a,N_\tau) \left<P\right>
     =-{\left.{\partial f_s}\over{\partial h}\right|}_{h=0}
 ~~,~~\EQNO{Pphys}
$$
$$
\chi_p={N_\tau}^d Z(a,N_\tau)^2 \chi
           ={\left.{\partial^2 f_s}\over{\partial h^2}\right|}_{h=0}
 ~~,~~\EQNO{chiphys}
$$
$$
g_4={\left.{\partial^4 f_s}\over{\partial h^4}\right|}_{h=0} \biggl/
    \left(\chi_p^2 {\left({N_\sigma \over N_\tau} \right)}^d\right)
 ~~.~~\EQNO{gr}
$$
\noindent
For the cumulant $g_4$ the renormalization factors cancel
and we end up with the usual expression known for the
Binder cumulant of the order parameter
\hbox{\hfil\refss{binder}{baker}{barberb}.\hfil}
The general form of the scaling relations derived from \eq{free} is
$$
\left.\partial^n f_s/\partial h^n \right|_{h=0} =
{\left(N_\sigma\over {N_\tau}\right)}^{n{{\beta+\gamma}\over \nu}-d} \cdot
Q_n\left(g_t\left({N_\sigma\over N_\tau}\right)^{1\over \nu}\right)
\EQNO{scale}
$$
\noindent
where the function $Q_n$ is defined as
$$
Q_n(x_1)={\left.{{\partial^n\over{{\partial x_2}^n}}
          Q_{f_s}(x_1,x_2)}\right|}_{x_2=0}.
\EQNO{Qn}
$$
The finite-size scaling behaviour of higher order cumulants
which are closely related to the derivatives defined in \eq{scale}
is discussed in appendix A. We note that for $g_4$, as well as the
higher cumulants, the prefactors of the scaling functions cancel.
Thus they take on unique fixed point values at $g_t = 0$ even on
finite lattices, if we ignore corrections from irrelevant operators.
The cumulants are thus well suited to determine the critical
coupling from simulations on finite lattices.

Let us first consider the case when $N_\tau$ is kept fixed. Then
$N_\tau$ can be absorbed in the non-universal constants in $g_t$
and $g_h$ and we end up with the usual form of the finite-size scaling
ansatz for $f_s$. The critical coupling can be determined from the
fixed point of $g_4(\beta,N_\sigma )$
\hbox{\hfil\refsss{engb}{binder}{barberb}{FerLan}.\hfil}

Next we consider $y=N_\sigma/N_\tau$ fixed, varying $N_\sigma$ and therefore
$N_\tau$ accordingly as is needed to reach the continuum limit.
Rescaling $N_\sigma$ and $N_\tau$ by a factor $b$ leads to a
phenomenological renormalization $\tilde\beta(\beta;b;y)$ by the following
identity for a scaling function $Q$
$$
Q(g_t(\beta,N_\tau)\left({N_\sigma\over N_\tau}\right)^{1\over \nu}) =
Q(g_t(\tilde\beta,b N_\tau) \left({bN_\sigma\over bN_\tau}\right)^{1\over \nu})
 ~~.~~  \EQNO{renorm}
$$
\noindent
Here we have kept explicit the dependence of $g_t$ on \nt that comes in
through the non-universal metric coefficients $c_t$ and $b_t$ in \eq{gt}.

The property that the normalized fourth cumulant
$g_4$ is directly a scaling function can be used to measure
the discrete $\beta$-function, $\Delta\beta$, by using the above identity
for $Q=g_4$. By writing \eq{renorm} in terms of the scaling
function $Q$ and not for the scaling field $g_t$ directly we do not have
to determine the metric coefficients in \eq{gt}.
The discrete $\beta$-function is given by the shift
in the coupling $\beta$, which is necessary to get the
same value of $g_4$ for the two different lattice sizes
$$
g_4(\beta-\Delta\beta_y(\beta);{N_\sigma};{N_\tau}) =
g_4(\beta;{bN_\sigma};{bN_\tau})  ~~.~~
\EQNO{renormgr}
$$
The function $\Delta\beta_y$ defined in \eq{renormgr} may depend on $y$
through contributions from irrelevant scaling fields
$g_i(\beta,N_\tau)N_\sigma^{y_i}$ with negative exponents $y_i$.
The discrete $\beta$-function can then be obtained by an
extrapolation to $y=\infty$
$$
\Delta\beta(\beta)=\lim_{y\to\infty} \Delta\beta_y(\beta) ~~.~~
\EQNO{limdb}
$$
The knowledge of $\beta_{c,\infty}(N_\tau )$ and $\Delta\beta$ allows
to calculate the critical coupling for the rescaled lattice size
$$
\beta_{c,\infty}(bN_\tau )=\beta_{c,\infty}(N_\tau ) + \Delta\beta ~~.~~
\EQNO{extrad}
$$
In the following we will use this approach to determine
$\beta_{c,\infty}$ for $N_\tau=8$ and 16. In particular
we will check the $y$-independence for $N_\tau\le 8$ and
use this information to justify our calculations for
$N_\tau=16$ with moderate values for $y$.

\bigskip
\noindent
\noindent
{\bf 3. Critical couplings and the Binder Cumulant
        for $\bf N_\tau =8$ and $\bf N_\tau =16$ }

As discussed in the previous section the Binder cumulants are well suited
to determine the critical couplings for the deconfinement transition.
Previously they have been used to determine the critical couplings for the
$SU(2)$ gauge theory on lattices with temporal extent $\nt =4$ and 6
\refs{enga}{engb}. We follow here the same strategy to determine the
critical coupling for $\nt =8$. We have performed simulations on lattices
of size $8 \times \ns ^3$ with $\ns =16$, 24 and 32 at several values of
$\beta$ in the vicinity of the estimated critical point. We have used an
overrelaxed heat-bath algorithm with 14 overrelaxation steps between
subsequent ``incomplete'' heat-bath updates \ref{marcu}. We used the
Kennedy-Pendleton \ref{kp} algorithm with one trial per link. The
acceptance rate was always larger than 96\%. Most runs were done on the
massively parallel CM-2 and so the use of a complete heat-bath algorithm
would have led to a considerable waste of resources. Measurements were
taken after each heat-bath update.
Details on our parameter choices
and number of iterations are given in \table{stat}, where we also give
results for the estimated integrated autocorrelation times for the
expectation value of the Polyakov loop.

\midinsert
$$\vbox{\offinterlineskip
\halign{
\strut\vrule     \hfil $#$ \hfil  &
      \vrule # & \hfil $#$ \hfil  &
      \vrule # & \hfil $#$ \hfil  &
      \vrule # & \hfil $#$ \hfil  &
      \vrule # & \hfil $#$ \hfil
      \vrule \cr
\noalign{\hrule}
  ~N_\sigma~
&&~N_\tau~
&&~\beta~
&&~N_{meas}~
&&~\tau_{int}~\cr
\noalign{\hrule}
 ~16~&&~~8~&&~2.500~&&~25010~&&~~3.4~\cr
 ~16~&&~~8~&&~2.510~&&~21055~&&~~3.7~\cr
 ~16~&&~~8~&&~2.520~&&~21200~&&~~3.6~\cr
\noalign{\hrule}
 ~24~&&~~8~&&~2.500~&&~55500~&&~10.6~\cr
 ~24~&&~~8~&&~2.505~&&~29500~&&~10.0~\cr
 ~24~&&~~8~&&~2.510~&&~29500~&&~13.5~\cr
 ~24~&&~~8~&&~2.515~&&~39500~&&~~9.8~\cr
 ~24~&&~~8~&&~2.520~&&~14500~&&~14.9~\cr
\noalign{\hrule}
 ~32~&&~~8~&&~2.500~&&~30354~&&~29.6~\cr
 ~32~&&~~8~&&~2.505~&&~34750~&&~17.1~\cr
 ~32~&&~~8~&&~2.510~&&~38540~&&~15.3~\cr
 ~32~&&~~8~&&~2.515~&&~28750~&&~18.0~\cr
 ~32~&&~~8~&&~2.520~&&~18800~&&~30.0~\cr
\noalign{\hrule}
 ~32~&&~16~&&~2.720~&&~30910~&&~10.9~\cr
 ~32~&&~16~&&~2.740~&&~40350~&&~12.6~\cr
 ~32~&&~16~&&~2.750~&&~35000~&&~12.0~\cr
\noalign{\hrule}
 ~48~&&~16~&&~2.740~&&~43850~&&~45.0~\cr
\noalign{\hrule}
}}$$
\centerline {\bf \table{stat}: \rm Run parameters.}
\bigskip
\endinsert
In \fig{g4_8} we show the Binder cumulant, which on a lattice is measured by
$$
g_4 = {\langle P^4 \rangle \over \langle P^2 \rangle ^2} - 3
 ~~,~~\EQNO{binder}
$$
with the Polyakov loop, $P$, given by \eq{pol}.
Also shown in this figure is an interpolation between results obtained at the
various values of $\beta$, which is based on the density of states method
(DSM) \ref{fer}.
Our implementation of the DSM takes into account each measurement and does
not require the usual division of the range of the expectation value of the
plaquette operator $U_p=N^{-1} {\rm Tr} U_1U_2U_3^{\dagger} U_4^{\dagger}$
in bins. This corresponds to an infinite number of bins and has the
advantage to remove a free parameter, the number of bins, from the method.
Because of the finite length of the Monte Carlo runs, the DSM provides
reliable results only if histograms of $U_p$ belonging to adjacent
couplings overlap. We convinced ourselves that at least 2.5\% of the data
of one sample is contained in each of the overlapping tails of the
distribution.

\if \preprint Y
\midinsert
\vskip3.5in
\includegraphics{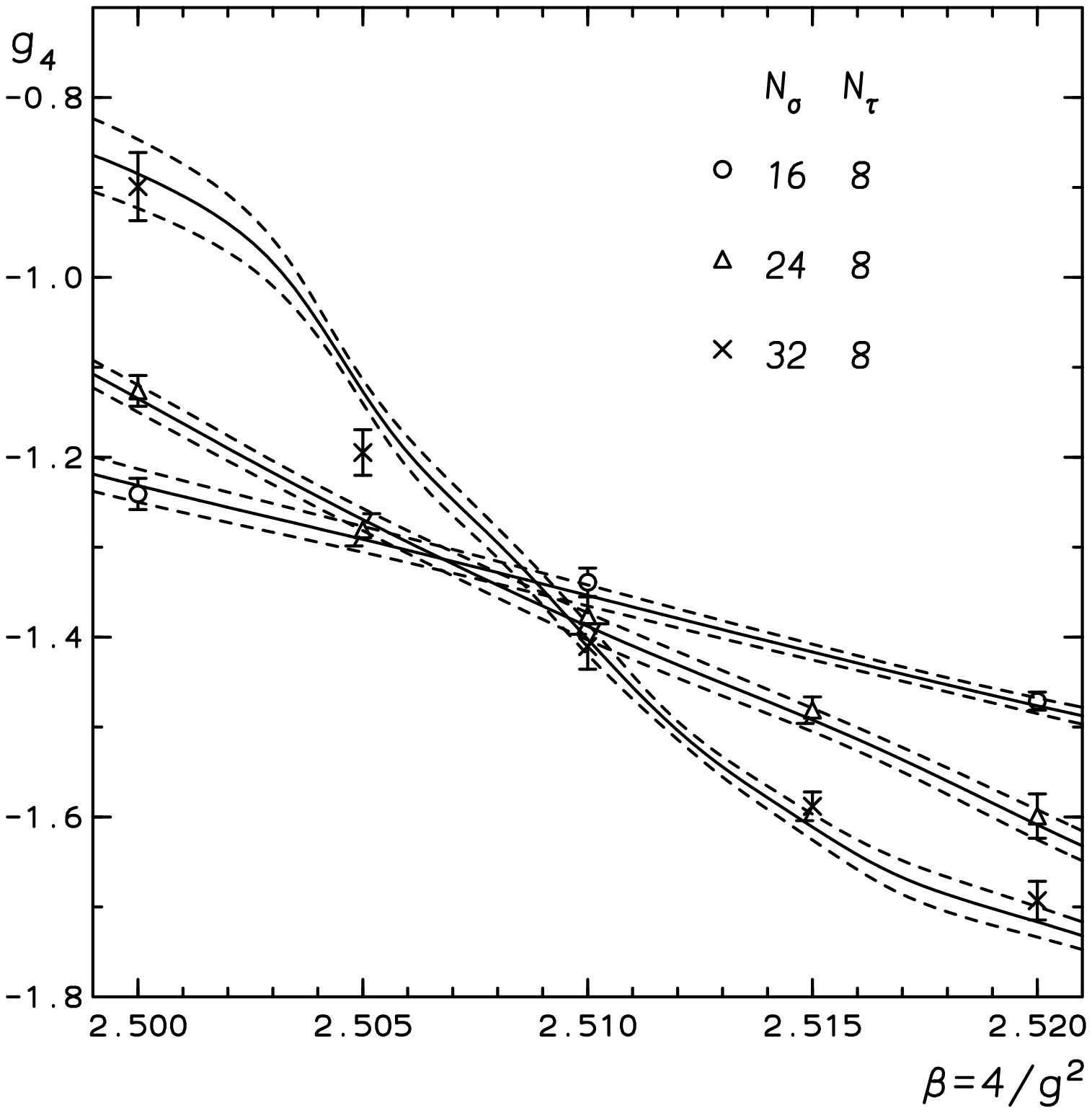}
\vskip.1in
\caption{\figtag{g4_8}. The cumulant \gr for \nta and various values of the
spatial lattice size as a function of the coupling $\beta$. Solid curves
are interpolation curves based on the density of states method. The dashed
lines indicate the error on the curves estimated by the jackknife method.}
\endinsert
\fi

The DSM allows an accurate determination of the intersection
points of $g_4(\beta;N_\sigma;N_\tau)$ and $g_4(\beta;bN_\sigma;N_\tau)$,
where $N_\sigma$ is the spatial size of the smaller lattice and
$b$ is given by the ratio ${{N_\sigma}^\prime} / N_\sigma$.
Taking also the largest irrelevant scaling field into account the
coupling $\beta_c(N_\sigma,b)$ at which two cumulants intersect
varies with $N_\sigma$ and $b$ as \refss{engb}{binder}{barberb}
$$\eqalign{
\beta_c(N_\sigma,b)&=\beta_{c,\infty} (1 - a \epsilon) \cr
\epsilon&=N_\sigma^{y_1-1/ \nu} {{1-b^{y_1}} \over {b^{1/ \nu}-1}}\cr
          }~~\lower1.45ex\hbox{.}
\EQNO{extrabc}
$$
Using the value $\nu=0.628$ calculated for the three-dimensional
Ising model \ref{FerLan} and $y_1=-1$ \ref{engb} we determine the
critical coupling for $\nt =8$ by an extrapolation to
$\epsilon=0$ with the result
$$
\beta_{c,\infty}(\nt =8)=2.5115 \pm 0.0040 ~~.~~ \EQNO{betaca}
$$
The error has been obtained from a weighted linear fit to
$\beta_c(N_\sigma,b)$, where the error for the intersection
points has been calculated using the jackknife method.
The extrapolation is shown in \fig{extra}.

\if \preprint Y
\midinsert
\vskip3.5in
\includegraphics{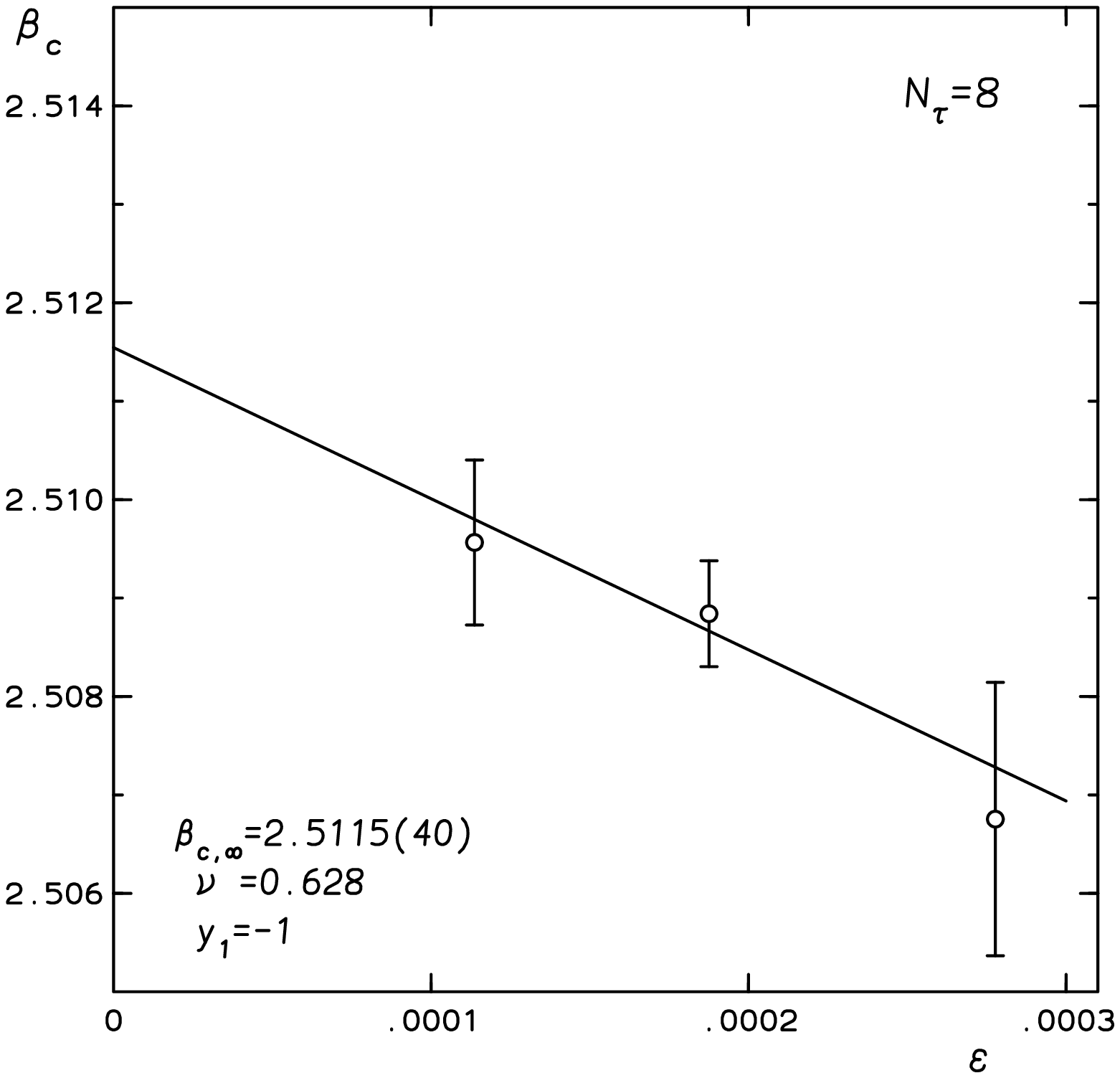}
\vskip.1in
\caption{\figtag{extra}. The coupling value at the intersection points of
$g_4$ for $N_\sigma=16,24$ and 32 as a function of $\epsilon$ which is
defined in \eq{extrabc}. The critical coupling $\beta_{c,\infty}$ can be
read from the figure as the section of the y axis at $\epsilon=0$.}
\endinsert
\fi

Taking into account the error band on the continuous curves
obtained by the density of states method shown in \fig{g4_8}
as well as the error on $\beta_{c,\infty}$ we obtain for
the fixed point value of $g_4$,
$$ \eqalignno{
\bar g_4(\nt )  &=\lim_{N_\sigma \to\infty}
                  g_4(\beta_{c,\infty};N_\sigma;N_\tau) & \EQNM{uni} \cr
\bar g_4(\nt =8)&=-1.48\pm 0.10 ~~.~~ & \EQNM{grfix} \cr
             }
$$
{}From the FSS theory as discussed in section 2 we expect,
in fact, that $\bar g_4$
is independent of $N_\tau$, \ie is
universal also in the continuum limit.
Indeed, this seems to be supported by our data.
In \table{g4fix} we give results
for $\bar g_4(\nt)$ obtained from simulations with $\nt =4$, 6 and 8.

\midinsert
$$
\vbox{\offinterlineskip
\halign{
\strut\vrule     \hfil $#$ \hfil  &
      \vrule # & \hfil $#$ \hfil
      \vrule \cr
\noalign{\hrule}
  ~N_\tau~
&&~\bar g_4~\cr
\noalign{\hrule}
 ~~~4~~~&&~~~-1.38(~3)~~~\cr
 ~~~6~~~&&~~~-1.43(~8)~~~\cr
 ~~~8~~~&&~~~-1.48(10)~~~\cr
\noalign{\hrule}}}
$$
\centerline{\bf \table{g4fix}: \rm Fixed point value of $\bar g_4$ for
different $N_\tau$.}
\bigskip
\endinsert
All three values agree reasonably well within errors,
although there seems to be a tendency to lower values
of $\bar g_4$ for larger $N_\tau$. At least partially this may be
related to still too-low statistics for lattices with
large $N_\tau$ as $g_4$ is a non-self-averaging
quantity \ref{FLB}. We note, however, that a common value of $\bar g_4$
taken from the spread of the data for $N_\tau=4,6,8$
and $16$ shown in \fig{g4univ} at
$t(N_\sigma/N_\tau)^{1/ \nu}=0$
shows good agreement with the value of the three-dimensional
Ising model \refs{FerLan}{privmanb}
$$\eqalignno{
\bar g_4              &=-1.40(10)       & \rm{\EQNM{grsu2}} \cr
\bar g_{4,{\rm Ising}}&=-1.41(~1) ~~.~~ & \rm{\EQNM{grising}} \cr
            }
$$

As discussed in section 2, $g_4$ is a scaling function, which in the continuum
limit depends only on the reduced temperature, $t = (T-T_c)/T_c$, and the
dimensionless quantity $y=LT$. In terms of lattice variables, $y$ is
given by
$$
y = {\ns \over \nt}~~,~~ \EQNO{y}
$$
and the reduced temperature in the asymptotic scaling regime is
given by \eq{t}. For small values of $t$ one expects,
$$
g_4(t,y) = g_4(ty^{1/\nu}) = \bar g_4 + g_{4,1} ty^{1/\nu}
 ~~.~~ \EQNO{guniv}
$$
This universal scaling behaviour indeed seems to be fulfilled quite well,
as can be seen from \fig{g4univ}, where we show $g_4(t,y)$ for various
values of \nt and $N_\sigma$. The critical exponent $\nu$ has been taken
from the three-dimensional Ising model \ref{FerLan}. Otherwise the
presentation in \fig{g4univ} is parameter-free.
We note that indeed the slope $g_{4,1}$ seems to be universal and
within our accuracy does not show any further
$N_{\tau}$ or even $N_{\sigma}$ dependence.

\if \preprint Y
\midinsert
\vskip3.5in
\includegraphics{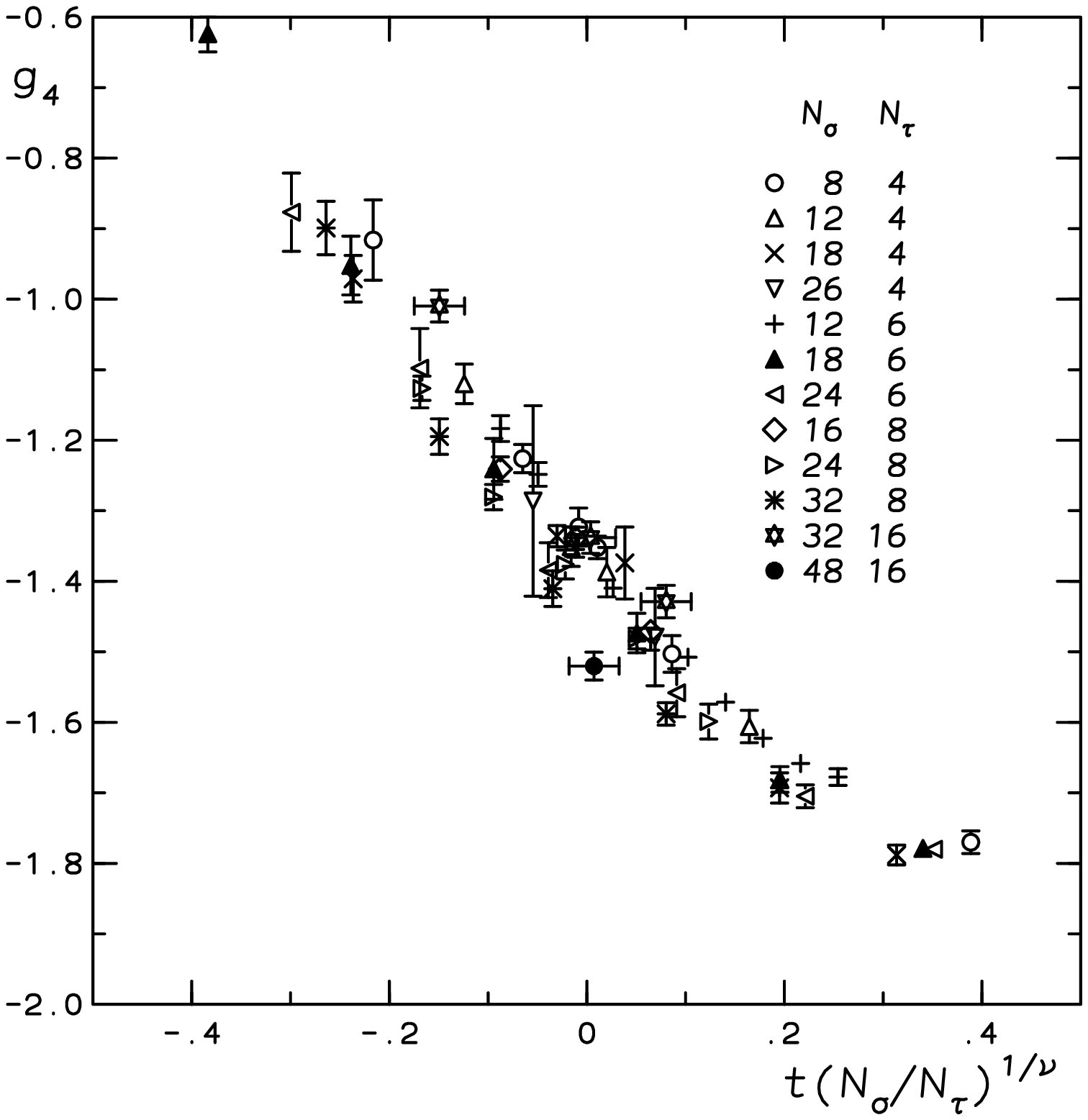}
\vskip.1in
\caption{\figtag{g4univ}. The Binder cumulant \gr as a function of
$ty^{1/\nu}$ for various lattice sizes as given in the figure. The critical
exponent $\nu$ has been taken to be the one of the three-dimensional Ising
model, $\nu =0. 628$. For $N_\tau=16$ and $N_\sigma=32$ and 48 we also mark
the error in x direction caused by the uncertainty in the critical coupling
$\beta_{c,\infty}$.}
\endinsert
\fi

The universal scaling behaviour of $g_4$ in the vicinity of the critical
point can be explored to determine the critical couplings for larger values
of \nt without an explicit determination of the fixed point $\bar g_4$ from
simulations on lattices of varying spatial size. We have calculated $g_4$
at three $\beta$ values on a $16 \times 32^3$ lattice and at one $\beta$
value on a $16 \times 48^3$ lattice. Details for these runs are also given
in \table{stat}. In \fig{g4_16} we show the Binder cumulant as a function
of $\beta$. The critical coupling $\beta_c(\nt =16)$ can now be determined
from the shift of $\beta$ required to overlay the data of \fig{g4_16} with
those for $N_\sigma$=16 and 24 shown in \fig{g4_8}. A matching procedure
according to \eq{renormgr} gives the shift values
$\Delta\beta_{y=2}=0.232(2)$ and $\Delta\beta_{y=3}=0.224(2)$.  The two
values show a noticeable difference, although the statistical error is
relatively large. To investigate whether the observed $y$ dependence is of
significance, we checked if this is present also for $N_\tau=4,6$ and $8$.
We determined the shift $\Delta\beta_y$ for $y=2,3$ and 4 using data from
\refs{enga}{engb} and found that there is no significant $y$ dependence of
$\Delta\beta_y$. Furthermore  $\Delta\beta_y$ agrees with the shift
$\Delta\beta=\beta_{c,\infty}(N_{\tau,1})-\beta_{c,\infty}(N_{\tau,2})$
calculated from the infinite volume critical coupling. For the largest
lattice size $\nt =16$ the critical slowing down is most severe, as is seen
from $\tau_{int}$  given in \table{stat}. As mentioned before $g_4$ is a
non-self-averaging quantity and, as for the susceptibility, the expectation
value will be too small, if one has not enough independent measurements
\ref{FLB}. This seems to be true for our dataset on the $N_\sigma =48$,
$N_\tau =16$ lattice and we can not obtain the same precision as in the
previous cases of $N_\tau=4,6$ and 8.

\if \preprint Y
\midinsert
\vskip3.5in
\includegraphics{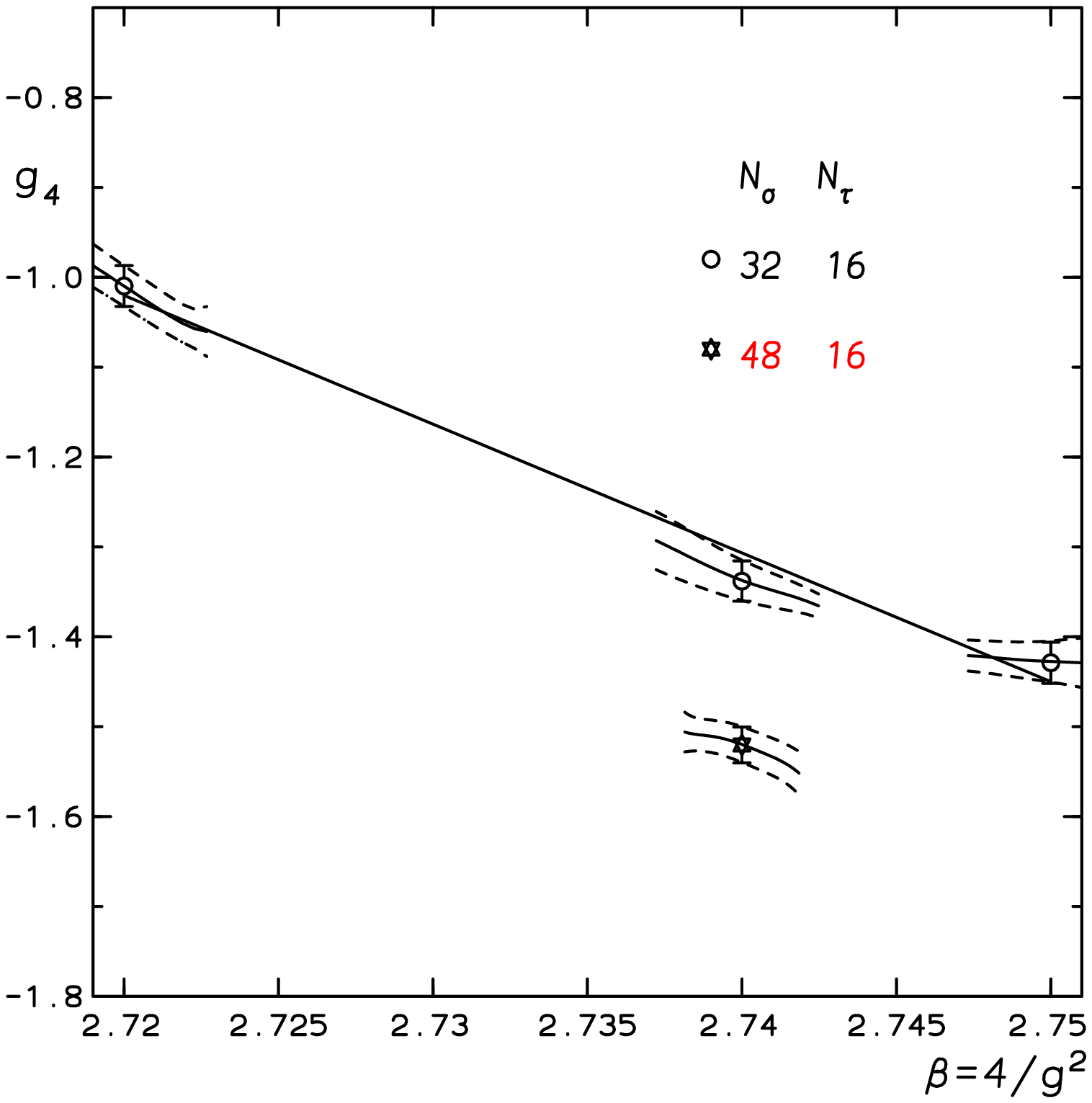}
\vskip.1in
\caption{\figtag{g4_16}. The cumulant $g_4$ for $N_\tau$ and $N_\sigma =
32$ (dots) and 48 (star) as a function of $\beta$. The straight line
represents a linear fit to the data points for $N_\sigma=32$. The four
curves through the data points are obtained using the single histogram
version of the DSM. The range of validity follows from the condition that
at least 2.5\% of the measurements in the data sample is contained in each
tail of the distribution of $U_p$. This gives the range of validity in
$U_P$, which is converted to a coupling range using the computed energy
coupling relation. The dashed lines indicate the error on these curves.}
\endinsert
\fi

Averaging over the two values of $\Delta\beta_y$ at $y=2$ and 3 and
assigning a large error, which includes both
numbers, we get $\Delta\beta(N_\tau=16)=0.228(6)$,
which is consistent with results obtained on
lattices of size $32^4$ \ref{defor}.
Using \eqand{extrad}{betaca} this corresponds to a critical coupling,
$$
\beta_{c,\infty}(\nt =16)=2.7395\pm 0.0100 ~~.~~ \EQNO{betacb}
$$

We want to point out that the combination of Monte Carlo simulation and the
DSM is especially suited to determine the intersection points of $g_4$ for
small volumes. If the system becomes too large, the widths of the histograms
of $U_p$ become very small and one needs a lot of simulations at different
$\beta$ values to cover a given range of the coupling. This was the case
for our results on lattices with $N_\tau=16$ and for these we could only
use the single histogram version of the DSM.

\noindent
{\bf 4. Scaling and Asymptotic Scaling}
\medskip
The critical couplings determined in the previous section can be used to
test scaling and asymptotic scaling in $SU(2)$ gauge theory over a much
wider range of the coupling than was previously possible. Additionally we
will compare the $SU(2)$ results to existing data for $SU(3)$ and four
flavour QCD. After that we will try to point out common features of the
data.

Recently the heavy quark potential has been studied in the pure $SU(2)$
gauge theory on quite large lattices and several values of the gauge
coupling \hbox{\hfil\refssss{pota}{potb}{mica}{mica1}{micb},\hfil}
which cover the range of couplings we have studied here.
The string tension extracted from the long distance part
of the heavy quark potential can be used to test the scaling of
dimensionless ratios of physical quantities. In particular we will
discuss here the scaling of $T_c/ \sqrt{\sigma}$. In \table{stringsu2}
and \table{stringsu3} we
have collected results from most recent calculations of the string
tension in $SU(2)$ \hbox{\hfil\refssss{pota}{potb}{mica}{mica1}{micb}\hfil}
as well as $SU(3)$ \refs{stra}{strb} gauge theory.
For four flavour QCD we used the value $\sqrt{\sigma}a=0.332(2)$ \ref{mtcb}.

\midinsert
$$\vbox{\offinterlineskip
\halign{
\strut\vrule     \hfil $#$ \hfil  &
      \vrule # & \hfil $#$ \hfil  &
      \vrule # & \hfil $#$ \hfil  &
      \vrule # & \hfil $#$ \hfil
      \vrule \cr
\noalign{\hrule}
  ~N_\sigma~
&&~N_\tau~
&&~\beta~
&&~\sqrt{\sigma}~a~\cr
\noalign{\hrule}
 ~~8~&&~10~&&~2.20~&&~0.4690(100)~\cr
 ~10~&&~10~&&~2.30~&&~0.3690(~30)~\cr
 ~16~&&~16~&&~2.40~&&~0.2660(~20)~\cr
 ~32~&&~32~&&~2.50~&&~0.1905(~~8)~\cr
 ~20~&&~20~&&~2.60~&&~0.1360(~40)~\cr
 ~32~&&~32~&&~2.70~&&~0.1015(~10)~\cr
 ~48~&&~56~&&~2.85~&&~0.0630(~30)~\cr
\noalign{\hrule}}}
$$
\centerline {\bf \table{stringsu2}: \rm $SU(2)$ string tension.}
\medskip
\endinsert

\midinsert
$$
\vbox{\offinterlineskip
\halign{
\strut\vrule     \hfil $#$ \hfil  &
      \vrule # & \hfil $#$ \hfil  &
      \vrule # & \hfil $#$ \hfil  &
      \vrule # & \hfil $#$ \hfil
      \vrule \cr
\noalign{\hrule}
  ~N_\sigma~
&&~N_\tau~
&&~\beta~
&&~\sqrt{\sigma}~a~\cr
\noalign{\hrule}
 ~16~&&~32~&&~5.60~&&~0.5295(~9)~\cr
 ~16~&&~24~&&~5.70~&&~0.4099(12)~\cr
 ~16~&&~24~&&~5.80~&&~0.3302(15)~\cr
 ~16~&&~24~&&~5.90~&&~0.2702(19)~\cr
 ~32~&&~32~&&~6.00~&&~0.2269(62)~\cr
 ~24~&&~32~&&~6.20~&&~0.1619(19)~\cr
 ~32~&&~32~&&~6.40~&&~0.1214(12)~\cr
 ~32~&&~32~&&~6.80~&&~0.0738(20)~\cr
\noalign{\hrule}}}
$$
\centerline {\bf \table{stringsu3}: \rm $SU(3)$ string tension.}
\bigskip
\endinsert
We have used a spline interpolation between these values to determine
the string tension $\sqrt{\sigma} a(\beta)$ at the critical couplings
of the deconfinement transition. These critical couplings are displayed
in \table{betacsu2} and in \table{betacsu3}
\hbox{\hfil\refss{columbia}{tsubuka}{kennedy}.\hfil}
\midinsert
$$
\vbox{\offinterlineskip
\halign{
\strut\vrule     \hfil $#$ \hfil  &
      \vrule # & \hfil $#$ \hfil  &
      \vrule # & \hfil $#$ \hfil  &
      \vrule # & \hfil $#$ \hfil  &
      \vrule # & \hfil $#$ \hfil
      \vrule \cr
\noalign{\hrule}
  ~N_\tau~
&&~\beta_c~
&&~T_c/ \Lambda_L~
&&~T_c/ \LMS~
&&~\left.T_c/ \LMS\right|_E~\cr
\noalign{\hrule}
 ~~~2~~&&~~~1.8800(~30)~~&&~~29.7(2)~~&&~~1.499(11)~~&&~~0.852(~6)~~\cr
 ~~~3~~&&~~~2.1768(~30)~~&&~~41.4(3)~~&&~~2.089(16)~~&&~~1.213(11)~~\cr
 ~~~4~~&&~~~2.2986(~~6)~~&&~~42.1(1)~~&&~~2.125(~3)~~&&~~1.313(~3)~~\cr
 ~~~5~~&&~~~2.3726(~45)~~&&~~40.6(5)~~&&~~2.047(23)~~&&~~1.360(21)~~\cr
 ~~~6~~&&~~~2.4265(~30)~~&&~~38.7(3)~~&&~~1.954(15)~~&&~~1.354(13)~~\cr
 ~~~8~~&&~~~2.5115(~40)~~&&~~36.0(4)~~&&~~1.815(18)~~&&~~1.325(16)~~\cr
 ~~16~~&&~~~2.7395(100)~~&&~~32.0(8)~~&&~~1.616(41)~~&&~~1.271(37)~~\cr
\noalign{\hrule}}}
$$
\centerline {\bf \table{betacsu2}: \rm Critical couplings for $SU(2)$.}
\medskip
\endinsert
\midinsert
$$
\vbox{\offinterlineskip
\halign{
\strut\vrule     \hfil $#$ \hfil  &
      \vrule # & \hfil $#$ \hfil  &
      \vrule # & \hfil $#$ \hfil  &
      \vrule # & \hfil $#$ \hfil  &
      \vrule # & \hfil $#$ \hfil  &
      \vrule # & \hfil $#$ \hfil
      \vrule \cr
\noalign{\hrule}
  ~N_\sigma~
&&~N_\tau~
&&~\beta_c~
&&~T_c/ \Lambda_L~
&&~T_c/ \LMS~
&&~\left.T_c/ \LMS\right|_E~\cr
\noalign{\hrule}
 ~12    ~&&~~3~&&~5.5500(100)~&&~85.70 \pm 0.96~&&~2.975 \pm 0.033~&&~1.217
\pm 0.038~\cr
 ~\infty~&&~~4~&&~5.6925(~~2)~&&~75.41 \pm 0.02~&&~2.618 \pm 0.001~&&~1.318
\pm 0.012~\cr
 ~\infty~&&~~6~&&~5.8941(~~5)~&&~63.05 \pm 0.04~&&~2.189 \pm 0.001~&&~1.299
\pm 0.002~\cr
 ~16    ~&&~~8~&&~6.0010(250)~&&~53.34 \pm 1.50~&&~1.851 \pm 0.052~&&~1.152
\pm 0.043~\cr
 ~16    ~&&~10~&&~6.1600(~70)~&&~51.05 \pm 0.40~&&~1.772 \pm 0.014~&&~1.155
\pm 0.012~\cr
 ~16    ~&&~12~&&~6.2680(120)~&&~48.05 \pm 0.65~&&~1.668 \pm 0.023~&&~1.110
\pm 0.017~\cr
 ~16    ~&&~14~&&~6.3830(100)~&&~46.90 \pm 0.53~&&~1.628 \pm 0.018~&&~1.111
\pm 0.016~\cr
 ~24    ~&&~16~&&~6.4500(500)~&&~46.27 \pm 2.50~&&~1.537 \pm 0.087~&&~1.064
\pm 0.069~\cr
\noalign{\hrule}}}
$$
\centerline {\bf \table{betacsu3}: \rm Critical couplings for $SU(3)$.}
\bigskip
\endinsert
\noindent
The resulting values for the ratio,
$$
{T_c \over \sqrt{\sigma}}=
 \left(\sqrt{\sigma}~a\left(\beta_c^\infty\left(N_\tau\right)\right)~N_\tau
 \right)^{-1} ~~,~~
\EQNO{Tcsig}
$$
are shown in \fig{scal} as a function of $aT_c \equiv 1/N_\tau$.
Also shown in this figure is a result for four flavour
QCD, which combines the measured critical coupling for $N_\tau =8$
($\beta_c=5.15 \pm 0.05)~\ref{mtca}$ and the result for the string
tension measured at $\beta=5.15$ on a $24 \times 16^3$ lattice
\ref{mtcb}.

\if \preprint Y
\midinsert
\vskip3.5in
\includegraphics{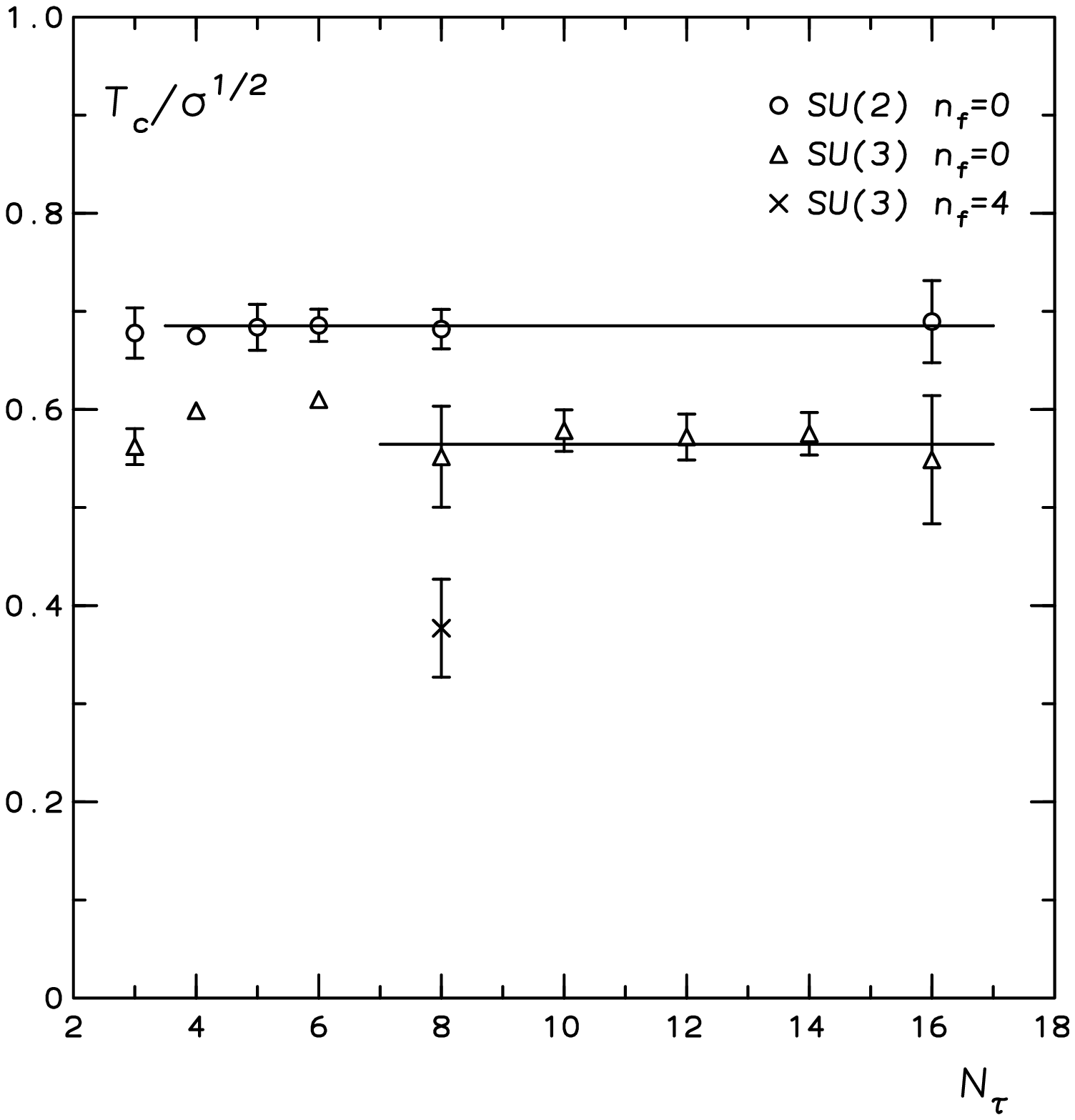}
\vskip.1in
\caption{\figtag{scal}. The critical temperature in units of the square
root of the string tension versus the $aT_c \equiv 1/N_\tau$ for the case
of $SU(2)$ (circles) and $SU(3)$ (triangles) pure gauge theory as well as
QCD with four flavours of dynamical fermions of mass $ma=0.01$ (cross). The
straight lines correspond to one parameter fits to the data.}
\endinsert
\fi

It is rather remarkable that scaling of $T_c / \sqrt{\sigma}$ is
valid in the case of $SU(2)$ to a high degree over the entire regime of
lattice spacings from $aT_c \le 0.25$ ($N_\tau \ge 4$) onwards.
For $SU(3)$ this seems to hold for $aT_c
\le 0.125$ ($ N_\tau \ge 8$), although for larger values of the lattice
spacing deviations are also only on the level of a few percent. Averaging
over the region $0.0625 \le aT_c \le 0.25$ ($4 \le N_\tau \le 16$) for
the $SU(2)$ data and over $0.0714 \le aT_c \le 0.125$
($8 \le N_\tau \le 14$) for $SU(3)$ we obtain
$$
{T_c \over \sqrt{\sigma}}= \cases{
0.69 \pm 0.02 &, SU(2) \cr
0.56 \pm 0.03 &, SU(3) \cr
0.38 \pm 0.05 &, QCD, $n_f=4$ \cr }
 ~~.~~\EQNO{ratiots}
$$
A strong decrease of this ratio with increasing number of degrees of
freedom (partons) is obvious. Such a behaviour is expected, as
a certain critical energy density can be reached already at a lower
temperature when the number of partons is larger. Moreover, it is
natural to assume that the string tension increases when the number of
gluonic degrees of freedom increases. Both effects tend to
lower the value of $T_c /\sqrt{\sigma}$ with increasing $N$ and/or $n_f$.

A more natural scale to compare the critical temperature in pure $SU(N)$
gauge theories is given by the glueball mass. The energy density in units
of $T^4$ of a massive, free glueball gas is a function of $x= m_G/T$ only,
$$
\epsilon /T^4 = d {x^3 \over 2 \pi^2} \sum_{n=1}^{\infty}
\bigl[ K_1(nx) +{3 \over nx} K_2(nx)
\bigr]
 ~~,~~\EQNO{gluegas}
$$
where $d$ is a degeneracy factor, $K_1$ and $K_2$ are modified
Bessel functions.
One thus might expect that the critical behaviour in the $SU(2)$ and $SU(3)$
theory is controlled by the ratio $T_c / m_G$,
with $m_G$ denoting the mass of the
lightest glueball. In fact, using recent glueball data
\hbox{\hfil\refss{mica}{baal}{micc},\hfil} we find that this
ratio varies only slightly between $SU(2)$ and $SU(3)$,
$$
{T_c \over m_G}= \cases{
0.180 \pm 0.016 &, SU(2) \cr
0.176 \pm 0.020 &, SU(3) \cr }
 ~~.~~\EQNO{ratiotg}
$$

It is well known that asymptotic scaling does not hold in the coupling
regime considered by us, despite the fact that scaling works remarkably
well. This suggests that there are universal scaling violating terms, which
are common to the different physical observables and thus may partially be
absorbed in a redefinition of the coupling constant. The need for such a
resummation has been noticed quite early and various schemes have been
suggested \hbox{\hfil\refssss{par}{fermilab}{samuel}{mak}{luescher}.\hfil}
In the case of the $SU(3)$ deconfinement
transition a variant of the scheme originally introduced by Parisi
\ref{par} has been found to yield quite good results already for rather
large lattice spacings \ref{pet}. Here the bare coupling, $\beta$, is
replaced by an effective coupling, $\beta_E$, which is related to the
plaquette expectation value and thus takes into account rapid fluctuations
in the pure gauge action at intermediate values of the coupling. Further
details on the definition of $\beta_E$ and a collection of plaquette
expectation values needed to calculate $\beta_E$ are given in appendix B.

In \fig{tcl} we show $T_c/\LMS$, using a conversion factor
$\LMS/\Lambda_L$ \ref{dashen} given by
$$
{\LMS\over \Lambda_L}=38.852704~\exp{\left(-{{3\pi^2}\over {11N^2}}\right)}
 ~~.~~ \EQNO{LMSL}
$$
\noindent
In the case of full QCD with four flavours we used the value
$\LMS/\Lambda_L = 76.45$ for the conversion factor.

\if \preprint Y
\midinsert
\vskip2.65in
\includegraphics{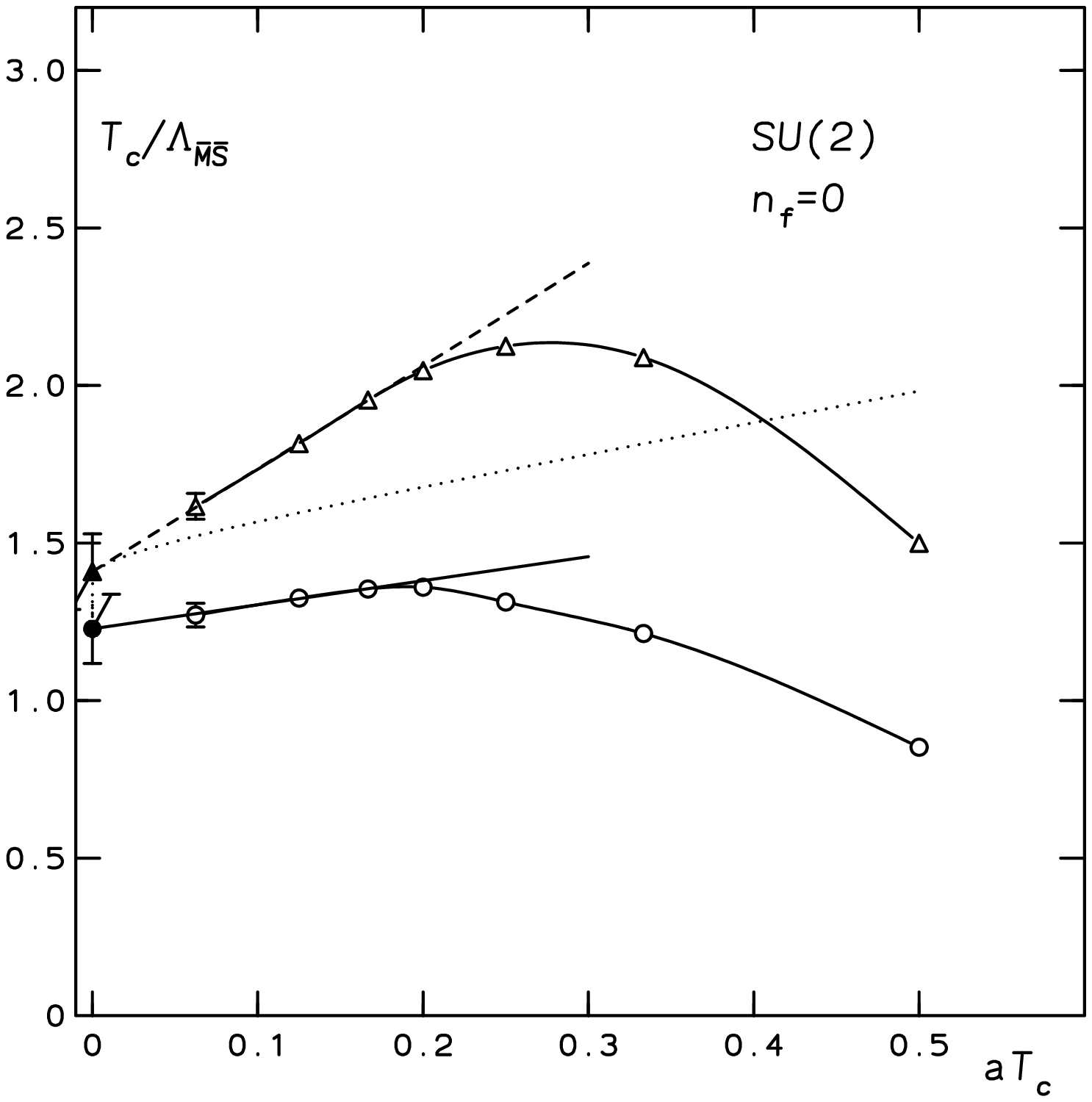}
\vskip2.65in
\includegraphics{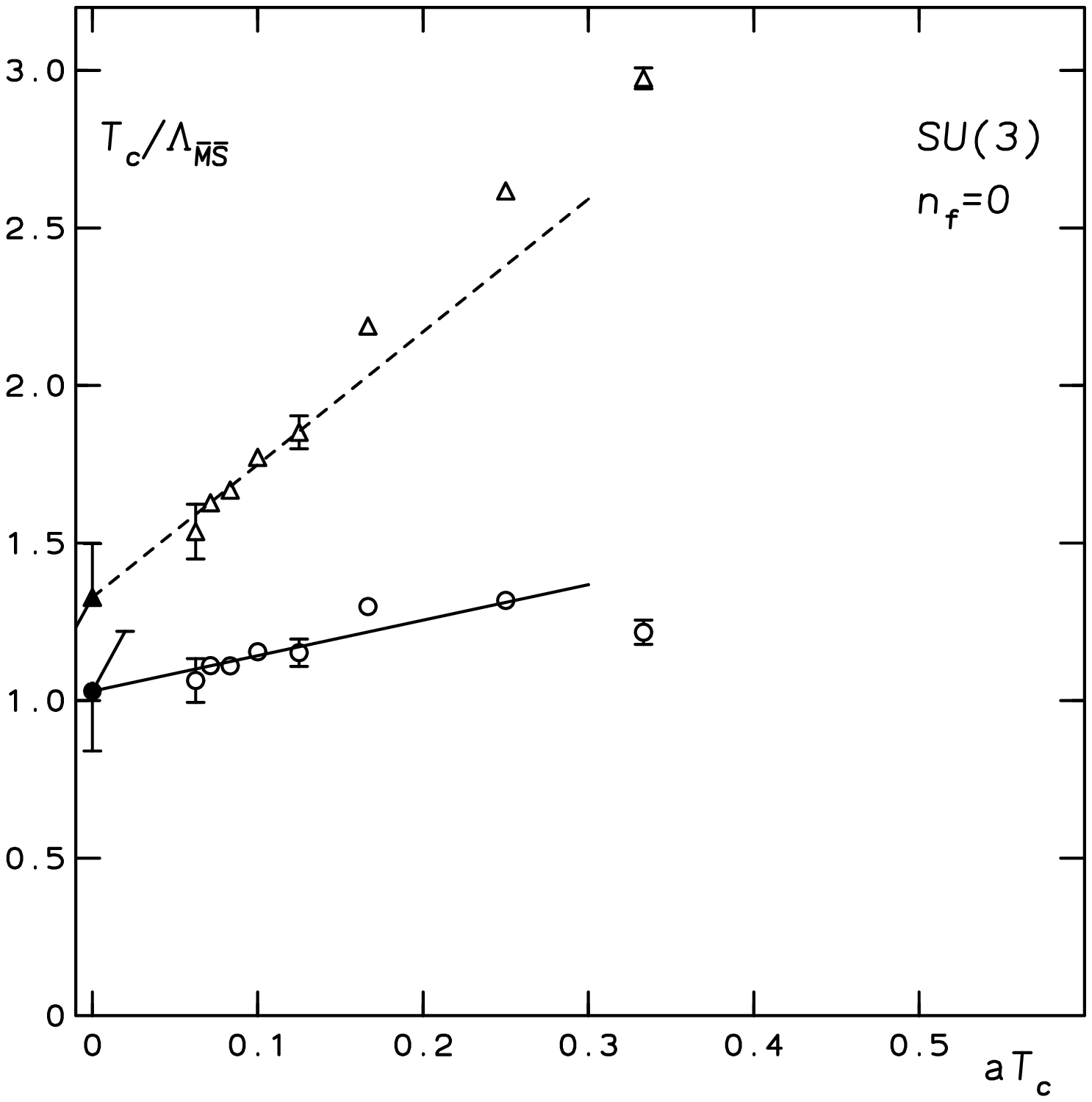}
\caption{\figtag{tcl}. The critical temperature in units of $\LMS$ for the
$SU(2)$ (a) and $SU(3)$ (b) gauge theory. Shown are data obtained by using
the bare coupling constant (triangles) and the effective coupling $\beta_E$
(circles) as input in the asymptotic renormalization group equation. The
solid straight lines give linear extrapolations of
these data sets to the continuum
limit, $a = 0$, in the effective coupling scheme. The broken lines indicate
a corresponding linear extrapolation using the bare coupling. The position
of the filled symbols marks the result of the extrapolation. The dotted
line in (a) marks how both coupling schemes approach in the continuum
limit, where we assumed a linear form of $T_c/\LMS$ as a function of $a$
given by the fit for the effective coupling scheme and used the third order
expansion of the internal energy given in appendix B, Eq. (B.1).}
\endinsert
\fi

It is evident that the asymptotic scaling violations are much reduced when
$\beta_E$ is used as a coupling constant. This is particularly true for the
case of $SU(3)$ shown in \fig{tcl}b. The $O(g^7)$ term in the $SU(N)$
$\beta$-function  would lead to violations of asymptotic scaling, which are
of $O(1/\rm{ln}a)$. Our data for $T_c / \LMS$ at finite values of the
cut-off suggest, however, that the deviations from asymptotic scaling are
well approximated by a correction term which is $O(a)$. This holds for both
coupling schemes, with the bare coupling $\beta$ and with the effective
coupling $\beta_E$. In the latter case, though, the coefficient of the
correction term is much smaller.
Performing a linear extrapolation of
$T_c/\LMS$ to $a = 0$ in the effective coupling scheme we find
$$
{T_c\over\LMS} = \cases{
1.23 \pm 0.11 &, SU(2) \cr
1.03 \pm 0.19 &, SU(3) \cr}
 ~~,~~\EQNO{tccont}
$$
where we used the region $0.0625 \le aT_c \le 0.1667$ ($6 \le N_\tau \le 16$)
for the $SU(2)$ data and $0.0714 \le aT_c \le 0.1250$ ($8 \le N_\tau \le 14$)
for $SU(3)$.
We want to note that this result does not disagree with a
linear extrapolation for the bare coupling scheme within errors.

It is remarkable that $T_c$ in units of $\LMS$ differs only by
less than 20\% between $SU(2)$ and $SU(3)$ and also the
corresponding value for QCD with four flavours,
$T_c/\LMS =1.05(20)$ \ref{mtca},
is surprisingly close to the $SU(3)$ value.

A corresponding linear extrapolation of the string tension
$\sqrt{\sigma}/\LMS$ in the region
$0.0625 \le aT_c \le 0.1667$ ($6 \le N_\tau \le 16$) for $SU(2)$ and
$0.0714 \le aT_c \le 0.1250$ ($8 \le N_\tau \le 14$) for $SU(3)$
to the continuum limit $a = 0$ gives the result
$$
{\sqrt{\sigma}\over\LMS}=\cases{
1.79 \pm 0.12 &, SU(2) \cr
1.75 \pm 0.20 &, SU(3) \cr}
 ~~.~~\EQNO{sigcont}
$$
The data from \table{stringsu3} together with the extrapolated values
are shown in \fig{sigl}.
The corresponding value for four flavour QCD is
$\sqrt{\sigma}/\LMS =2.78 \pm 0.50$.
This result is, of course, consistent with the scaling value of
$T_c/\sqrt{\sigma}$ given in \eq{ratiots}.

\if \preprint Y
\midinsert
\vskip3in
\includegraphics{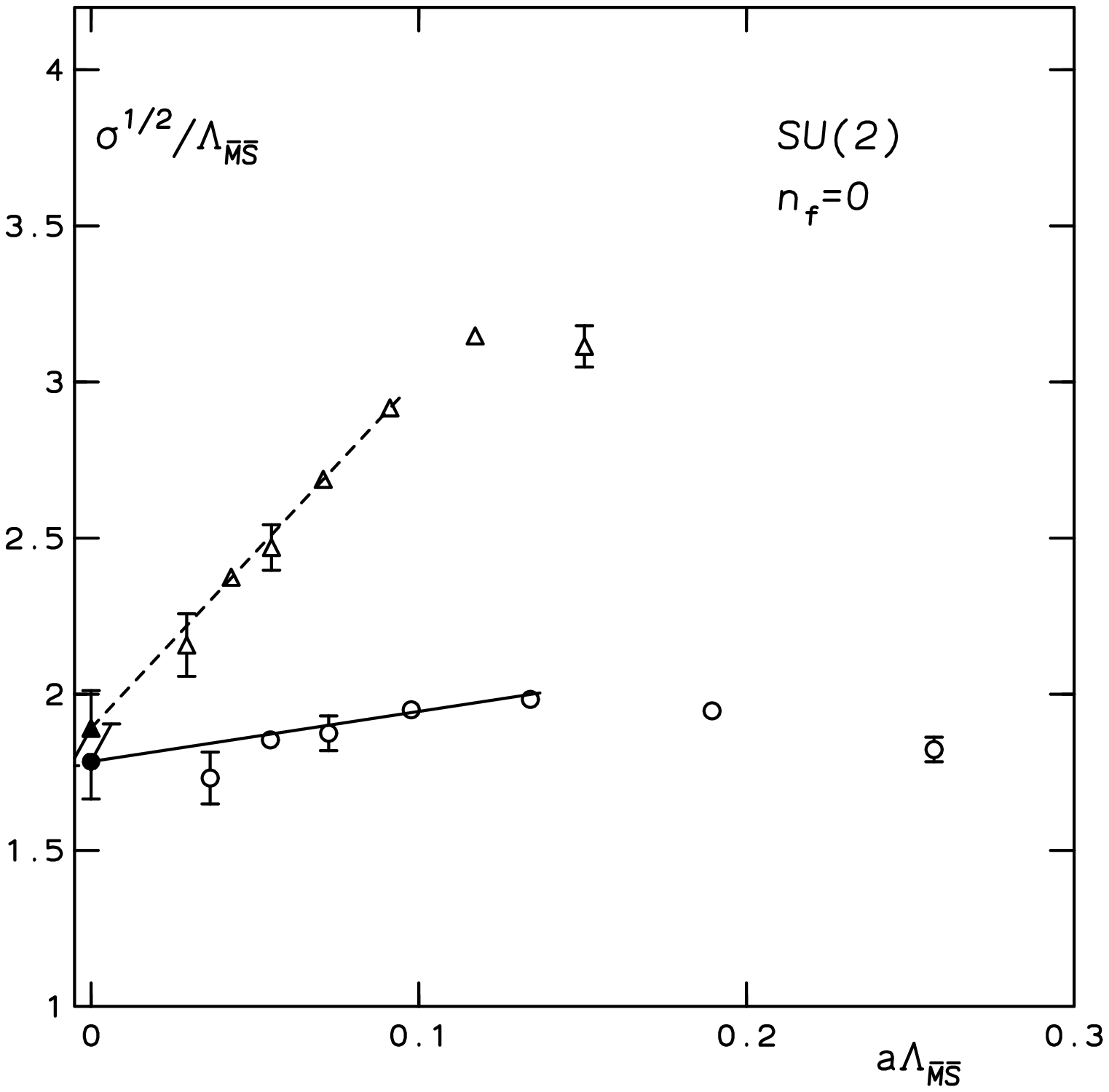}
\vskip.1in
\vskip3in
\includegraphics{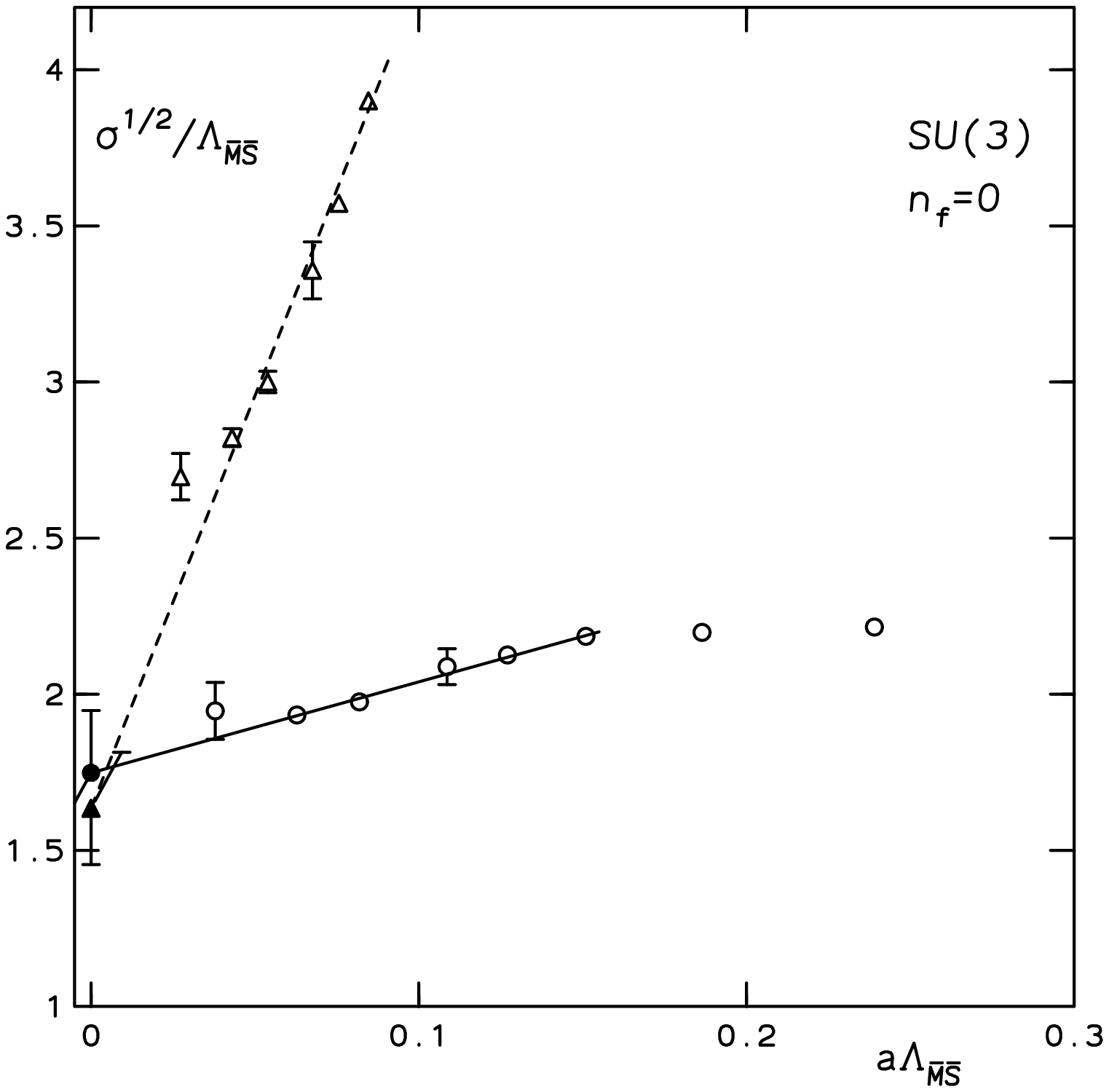}
\vskip.1in
\caption{\figtag{sigl}. The string tension $\sqrt{\sigma}$ in units of
$\LMS$ as a function of $a\LMS$ obtained from the two loop renormalization
group equation \eq{alam} for the $SU(2)$ (a) and $SU(3)$ (b) gauge theory.
The convention for symbols and lines is the same as in \fig{tcl}.}
\endinsert
\fi

We note that our value of the $SU(2)$ string tension
is only slightly larger than a recent estimate
from the heavy quark potential \ref{potb}, which
yielded $\sqrt{\sigma}/\LMS =1.61(9)$. The agreement between
these entirely different approaches gives additional
support for our $a = 0$ extrapolation in the
$\beta_E$-scheme.

\bigskip
\noindent
{\bf 5. Conclusions}
\medskip
We determined the critical coupling for the $SU(2)$
deconfinement transition on lattices with
large temporal extent, $N_\tau=8$ and $16$, using a
FSS analysis of the Binder cumulant $g_4$.
We find
$$\eqalign{
\beta_{c,\infty}(N_\tau = 8) &= 2.5115( 40) ~~,~~ \cr
\beta_{c,\infty}(N_\tau =16) &= 2.7395(100) ~~.~~ \cr
          }
$$
A comparison with existing string tension data showed
scaling in the entire coupling range $2.30 \le \beta \le 2.74$
under consideration. For the $SU(3)$ gauge group scaling sets in
at a value of $\beta\simeq 6$. We observe a 20\% increase in
$T_c/\sqrt{\sigma}$ when going from $SU(3)$ to $SU(2)$, which can
be understood in terms of the larger number of gluons in the
$SU(3)$ gauge theory, which tend to decrease $T_c$ and increase
$\sqrt{\sigma}$ at the same time.

In the bare coupling $\beta=2N/g^2$ we observe no sign of asymptotic
scaling up to $\beta=2.74$. The alternative coupling scheme $\beta_E$,
derived from the measured plaquette expectation value, shows much less
deviation from asymptotic scaling. Our data indicate that these
deviations are well described by $O(a)$ corrections and
a linear extrapolation of $T_c/ \Lambda_L$
and $\sqrt{\sigma}/ \Lambda_L$ to $a = 0$ seems to be justified.

We note, however, that our analysis clearly is not sensitive
to $O(1/\rm{ln}a)$ correction terms, which would be nearly constant
in the rather small interval of couplings considered by us.

\bigskip
\bigskip
\vbox{
\noindent
{\bf ACKNOWLEDGEMENT}
\bigskip
Our Monte Carlo simulations have been performed on the Connection
Machines CM-2 of the HLRZ-J\"ulich at GMD, Birlingshoven, of SCRI,
Tallahassee, on the Intel iPSC/860 at the ZAM, Forschungszentrum J\"ulich,
and on the NEC SX-3 computer of the RRZ K\"oln.
}
We thank these computer centers for providing us with the necessary
computational resources and the staff for their continuous
support. In particular we thank Dr. V\"olpel, GMD, for making the
CM-2 available to us during the Christmas break.
Financial support from DFG under contract \hbox{Pe~340/1-3} and from the
Ministerium f\"ur Wissenschaft und Forschung NRW under contract
\hbox{IVA5-10600990} is gratefully acknowledged (JF and FK).
The work of UMH was supported in part by the DOE
under grant \hbox{\# DE-FG05-85ER\discretionary{}{}{}250000}.
He also acknowledges partial
support by the NSF under grant \hbox{\# INT-8922411} and the kind hospitality
at the Fakult\"at f\"ur Physik at
the University of Bielefeld while this research was begun.

\vfill\eject
\bigskip\centerline{\bf REFERENCES}\bigskip
\item{\reftag{creutz})}
    M. Creutz, \PR {\bf D21} (1980) 2308.
\item{\reftag{wolff})}
    U.~Wolff, \PL {\bf 222B} (1989) 473; \NP {\bf B334} (1990) 581.
\item{\reftag{mcrga})}
    A.~Hasenfratz, P.~Hasenfratz, U.~Heller and F.~Karsch, \PL {\bf 140B}
                                                           (1984) 76.
\item{\reftag{mcrgb})}
    R.~Gupta, G.~W.~Kilcup, A.~Patel and S.~R.~Sharpe, \PL {\bf 211B}
                                                       (1988) 132.
\item{\reftag{pota})}
    S.~P.~Booth et al. (the UKQCD Collaboration), \PL {\bf 275B} (1992) 424.
\item{\reftag{potb})}
    C.~Michael, \PL {\bf 283B} (1992) 103.
\item{\reftag{enga})}
    J.~Engels, J.~Fingberg, and M.~Weber, \NP {\bf B332} (1990) 737.
\item{\reftag{engb})}
    J.~Engels, J.~Fingberg and D.~Miller, to appear in \NP B (1992).
\item{\reftag{barbera})}
    M.~N.~Barber, in Phase Transitions and Critical Phenomena Vol.~8,
    ed. C.~Domb and J.~Lebowitz, Academic Press (1981).
\item{\reftag{privmana})}
    V.~Privman, in Finite-Size Scaling and Numerical Simulations of
    Statistical Systems, World Scientific Publishing Co. (1990).
\item{\reftag{binder})}
    K.~Binder, \ZP {\bf B43} (1981) 119; \hfil\break
    K.~Binder, \PRL {\bf 47} (1981) 693.
\item{\reftag{baker})}
    G.~A.~Baker and J.~Kincaid, J. Stat. Phys. {\bf 24} (1981) 469; \hfil\break
    G.~A.~Baker and B.~A.~Freedman, J. Phys. {\bf A15} (1982) L715.
\item{\reftag{barberb})}
    M.~N.~Barber, R.~B.~Pearson, D.~Toussaint and J.~L.~Richardson,
    \PR {\bf B32} (1985) 1720.
\item{\reftag{FerLan})}
    A.~M.~Ferrenberg and D.~P.~Landau, \PR {\bf B44} (1991) 5081.
\item{\reftag{marcu})}
    K.~Fredenhagen and M.~Marcu, \PL {\bf 193B} (1987) 486.
\item{\reftag{kp})}
    A.~Kennedy and B.~Pendleton, \PL {\bf 156B} (1985) 393.
\item{\reftag{fer})}
    M.~Falconi, E.~Marinari, M.~L.~Paciello, G.~Parisi and
                             B.~Taglienti, \PL {\bf 108B} (1982) 331;
                             \hfil\break
    E.~Marinari, \NP {\bf B235} (1984) 123; \hfil\break
    G.~Bhanot, S.~Black, P.~Carter and
                         R.~Salvador, \PL {\bf 183B} (1986) 331; \hfil\break
    G.~Bhanot, K.~Bitar, S.~Black, P.~Carter and
                         R.~Salvador, \PL {\bf 187B} (1987) 381; \hfil\break
    G.~Bhanot, K.~Bitar and R.~Salvador, \PL {\bf 188B} (1987) 246; \hfil\break
    A.~M.~Ferrenberg and R.~H.~Swendsen, \PRL {\bf 61} (1988) 2635;
                                         \PRL {\bf 63} (1989) 1195.
\item{\reftag{FLB})}
    A.~M.~Ferrenberg, D.~P.~Landau and K.~Binder, J. Stat. Phys. {\bf 63}
                                                  (1991) 867.
\item{\reftag{privmanb})}
    V.~Privman, P.~Hohenberg and A.~Aharony, in Phase Transitions and Critical
    Phenomena Vol. 14, ed. C.~Domb and J.~Lebowitz, Academic Press (1991).
\item{\reftag{defor})}
    K.~M.~Decker and Ph.~de~Forcrand, \NP B (Proc. Suppl.) {\bf 17} (1990) 567.
\item{\reftag{mica})}
    C.~Michael and M.~Teper, \PL {\bf 199B} (1987) 95.
\item{\reftag{mica1})}
    S.~Perantonis, A.~Huntley and C.~Michael, \NP {\bf B326} (1989) 544;
                                              \hfil\break
    S.~Perantonis and C.~Michael, \NP {\bf B347} (1990) 854; \hfil\break
    C.~Michael and S.~Perantonis, \NP B (Proc. Suppl.) (1991) 177.
\item{\reftag{micb})}
    S.~P.~Booth et al. (The UKQCD Collaboration),
    Liverpool preprint LTH271 October 1991, LTH284 August 1992.
\item{\reftag{stra})}
    D.~Barkai, K.~J.~M.~Moriarty and C.~Rebbi, \PR {\bf D30} (1984) 1293;
                                               \hfil\break
    K.~C.~Bowler, F.~Gutbrod, P.~Hasenfratz, U.~Heller, F.~Karsch,
    R.~D.~Kenway, I.~Montvay, G.~S.~Pawley, J.~Smit and D.~J.~Wallace,
    \PL {\bf 163B} (1985) 367.
\item{\reftag{strb})}
    G.~S.~Bali and K.~Schilling, Wuppertal preprint WUB 92-02.
\item{\reftag{mtcb})}
    K.~D.~Born, R.~Altmeyer, W.~Ibes, E.~Laermann, R.~Sommer, T.~F.Walsh
    and P.~Zerwas (The $MTc$ Collaboration), \NP B (Proc. Suppl.) {\bf 20}
                                             (1991) 394.
\item{\reftag{columbia})}
    N.~H.~Christ, \NP B (Proc. Suppl.) {\bf 17} (1990) 267; \hfil\break
    H.~Ding and N.~Christ, \PR Lett. {\bf 60} (1988) 1367; \hfil\break
    F.~Brown et al., \PR Lett. {\bf 61} (1988) 2058.
\item{\reftag{tsubuka})}
    Y.~Iwasaki, K.~Kanaya, T.~Yoshi\'e, T.~Hoshino, T.~Shirakawa, \hfil\break
    Y.~Oyanagi, S.~Ichii and T.~Kawai, \PR Lett. {\bf 67} (1991) 3343;
                                       \hfil\break
    Y.~Iwasaki et al., University of Tsubuka preprint UTHE-237 (1992).
\item{\reftag{kennedy})}
    A.~D.~Kennedy, J.~Kuti, S.~Meyer and B.~J.~Pendelton,
    \PR Lett. {\bf 54} (1985) 87; \hfil\break
    S.~A.~Gottlieb, J.~Kuti, D.~Toussaint, A.~D.~Kennedy, S.~Meyer,
    B.~J.~Pendelton and R.~L.~Sugar, \PR Lett. {\bf 55} (1985) 1958.
\item{\reftag{mtca})}
    R.~V.~Gavai, S.~Gupta, A.~Irb\"ack, F.~Karsch, S.~Meyer,
    B.~Petersson, H.~Satz and H.~W.~Wyld (The $MT_c$ Collaboration),
    \PL {\bf 241B} (1990) 567.
\item{\reftag{baal})}
    P.~van~Baal and A.~S.~Kronfeld, \NP B (Proc.Suppl.) {\bf 9} (1990) 227.
\item{\reftag{micc})}
    C.~Michael and M.~Teper, \NP {\bf B314} (1989) 347; \hfil\break
    C.~Michael, \NP B (Proc. Suppl.) {\bf 17} (1990) 59.
\item{\reftag{par})}
    G.~Parisi, Proceedings of the $XX^{th}$ Conference on High Energy
    Physics, Madison 1980.
\item{\reftag{fermilab})}
    G.~P.~Lepage and P.~B.~Mackenzie,
    \NP B (Proc. Suppl.) {\bf 20} (1991) 173; \hfil\break
    A.~X.~El-Khadra, G.~Hockney, A.~Kronfeld and P.~B.~Mackenzie,
    FERMILAB-PUB-91/354-T (1991).
\item{\reftag{samuel})}
    F.~Green and S.~Samuel, \NP {\bf B194} (1982) 107;\hfil\break
    S.~Samuel, O.~Martin and K.~Moriarty, \PL {\bf 152B} (1984) 87.
\item{\reftag{mak})}
    Y.~M.~Makeenko and M.~I.~Polykarpov, \NP {\bf B205} (1982) 386.
\item{\reftag{luescher})}
    M.~L\"uscher, R.~Sommer and U.~Wolff, Cern preprint CERN-TH 6566/92 (1992).
\item{\reftag{pet})}
    F.~Karsch and R.~Petronzio, \PL {\bf 153B} (1985) 87.
\item{\reftag{dashen})}
    R.~Dashen and D.~J.~Gross, \PR {\bf D23} (1981) 2340.
\item{\reftag{heller})}
    U.~Heller and F.~Karsch, \NP {\bf B258} (1985) 29.
\item{\reftag{pana})}
    H.~Panagopoulos, private communication.
\item{\reftag{has})}
    F.~Gutbrod, P.~Hasenfratz, Z.~Kunszt and I.~Montvay,
    \PL {\bf 128B} (1983) 415; \hfil\break
    A.~Hasenfratz, P.~Hasenfratz, U.~Heller and F.~Karsch,
    \PL {\bf 143B} (1984) 193; \hfil\break
    K.~C.~Bowler, A.~Hasenfratz, P.~Hasenfratz, U.~Heller, F.~Karsch,
    R.~D.~Kenway, H.~Meyer-Ortmanns, I.~Montvay, G.~S.~Pawley,
    and D.~J.~Wallace, \NP {\bf B257} (1985) 155.
\item{\reftag{attig})}
    N.~Attig, PhD thesis, University of Bielefeld (1988).
\item{\reftag{bali})}
    G.~S.~Bali and K.~Schilling, Wuppertal preprint WUB 92-29.
\item{\reftag{laut})}
    M.~Creutz, L.~Jacobs and C.~Rebbi, \PRL {\bf 42} (1979) 1390;
                                       \PR {\bf D20} (1979) 1915;\hfil\break
    M.~Creutz, \PRL {\bf 43} (1979) 553;
    B.~Lautrup and M.~Nauenberg, \PR Lett. {\bf 45} (1980) 1755.
\item{\reftag{BC})}
    G.~Bhanot and M.~Creutz, \PR {\bf D24} (1981) 3212.
\item{\reftag{columbiab})}
    N.~Christ and A.~Terrano, Columbia preprint CU-TP-266.
\vfill\eject
\if \preprint N
\bigskip\centerline{\bf FIGURES}\bigskip
\item{Fig.~\figtag{g4_8}}
  The cumulant \gr for \nta and various values of the spatial lattice
  size as a function of the coupling $\beta$. Solid curves are
  interpolation curves based on the density of states
  method. The dashed lines indicate the error on the curves estimated
  by the jackknife method.
\vskip 10pt
\item{Fig.~\figtag{extra}}
  The coupling value at the intersection points of $g_4$ for
  $N_\sigma=16,24$ and 32 as a function of $\epsilon$ which is
  defined in \eq{extrabc}. The critical coupling $\beta_{c,\infty}$
  can be read from the figure as the section of the y axis at
  $\epsilon=0$.
\vskip 10pt
\item{Fig.~\figtag{g4univ}}
  The Binder cumulant \gr as a function of $ty^{1/\nu}$ for various
  lattice sizes as given in the figure. The critical exponent $\nu$ has
  been taken to be the one of the three-dimensional Ising model, $\nu =0.628$.
  For $N_\tau=16$ and $N_\sigma=32$ and 48 we also mark the error in
  x direction caused by the uncertainty in the critical coupling
  $\beta_{c,\infty}$.
\vskip 10pt
\item{Fig.~\figtag{g4_16}}
  The cumulant $g_4$ for $N_\tau$ and $N_\sigma = 32$ (dots) and 48 (star)
  as a function of $\beta$. The straight line represents a linear fit
  to the data points for $N_\sigma=32$. The four curves through the
  data points are obtained using the single histogram version of the
  DSM. The range of validity follows from the condition that at least
  2.5\% of the measurements in the data sample is contained in each tail
  of the distribution of $U_p$. This gives the range of validity
  in $U_P$, which is converted to a coupling range using the computed
  energy coupling relation. The dashed lines indicate the error on these
  curves.
\vskip 10pt
\item{Fig.~\figtag{scal}}
  The critical temperature in units of the square root of the string
  tension versus the $aT_c \equiv 1/N_\tau$ for the case of $SU(2)$ (circles)
  and $SU(3)$ (triangles) pure gauge theory as well as QCD with four
  flavours of dynamical fermions of mass $ma=0.01$ (cross). The straight
  lines correspond to one parameter fits to the data.
\vskip 10pt
\item{Fig.~\figtag{tcl}}
  The critical temperature in units of $\LMS$ for
  the $SU(2)$ (a) and $SU(3)$ (b) gauge theory.
  Shown are data obtained by using the bare coupling constant
  (triangles) and the effective coupling $\beta_E$ (circles) as input in the
  asymptotic renormalization group equation.
  The solid straight lines give linear
  extrapolations of these data sets to the continuum limit, $a = 0$,
  in the effective coupling scheme. The broken lines indicate a
  corresponding linear extrapolation using the bare coupling.
  The position of the filled symbols marks the result of the
  extrapolation.
  The dotted line in (a) marks how both coupling schemes approach
  in the continuum limit, where we assumed a linear form of
  $T_c/\LMS$ as a function of $a$ given by the fit for the effective
  coupling scheme and used the third order
  expansion of the internal energy given in appendix B, Eq. (B.1).
\vskip 10pt
\item{Fig.~\figtag{sigl}}
  The string tension $\sqrt{\sigma}$ in units of $\LMS$
  as a function of $a\LMS$ obtained from the two loop
  renormalization group equation \eq{alam} for the $SU(2)$ (a) and
  $SU(3)$ (b) gauge theory. The convention for symbols and
  lines is the same as in \fig{tcl}.

\vfil\eject
\fi

\noindent
{\bf Appendix A: Cumulants of the order parameter}
\medskip

We define cumulants of the Polyakov loop expectation value in the following
way:
$$ \eqalignno{
K_n &={\left(N_\sigma \over N_\tau\right)}^{d(1-n)}
      {\left.{\partial^n f_s}\over{{\partial h}^n}\right|}_{h=0} & \rm{(A.1)}
             }
$$

On a finite lattice all odd cumulants are zero due to the $Z(N)$ center
symmetry.
$$ \eqalignno{
   K_i &= 0,\rm{~~i~odd} & \rm{(A.2)} \cr
   K_2 &= <P^2>          & \rm{(A.3)} \cr
   K_4 &= <P^4> - 3<P^2>^2 & \rm{(A.4)} \cr
   K_6 &= <P^6> - 15<P^2><P^4> + 30<P^2>^3 & \rm{(A.5)} \cr
   K_8 &= <P^8> - 35<P^4>^2 - 28<P^2><P^6> + &~ \cr
   ~~~~&~~~~~420<P^2>^2<P^4> - 630<P^2>^3 & \rm{(A.6)} \cr}
$$

Then moments of the order parameter can be expressed by cumulants:
$$ \eqalignno{
   <P^2> &= K_2 & \rm{(A.7)} \cr
   <P^4> &= K_4 + 3K_2^2    & \rm{(A.8)} \cr
   <P^6> &= K_6 + 15K_2K_4 + 75K_2^3 & \rm{(A.9)} \cr
   <P^8> &= K_8 - 28K_6K_2 - 35K_4^4 - 210K_4K_2^2 - 1785K_2^4
 & \rm{(A.10)} \cr}
$$

Now we consider three possible cumulant ratios given by:
$$ \eqalignno{
   g_4&={{K_4}\over {K_2^2}}
       ={{<P^4>}\over{<P^2>^2}}-3 & \rm{(A.11)} \cr
   g_6&={{K_6}\over{K_2^3}}
       ={{<P^6>}\over {<P^2>^3}}-15g_4-75 & \rm{(A.12)} \cr
   g_8&={{K_8}\over{K_2^4}}
       ={{<P^8>}\over {<P^2>^4}}+28g_6+35g_4^2-210g_4-315 & \rm{(A.13)} \cr}
$$

The general form of these cumulant ratios obtained from Eq. (A.1) is
$$ \eqalignno{
   g_{n m} &= {K_{n m} \over K_n^m}
            = {\left({{N_\sigma}\over{N_\tau}}\right)^{d(1-nm)} \over
               \left({{N_\sigma}\over{N_\tau}}\right)^{dm(1-n)}}
      {{\left.\partial^{nm}f_s/\partial h^{nm}\right|_{h=0}} \over
 {\left(\left.\partial^n   f_s/\partial h^n   \right|_{h=0}\right)^m}}
 = {Q_{n m} \over Q_n^m} & \rm{(A.14)} \cr }
$$

The ratio $K_{nm}/K_n^m$ is directly a scaling function. Therefore it
can be expressed as a function of $g_4$ since both are derived from $f_s$
and depend on the same argument
$g_t \left({N_\sigma /N_\tau}\right)^{1/\nu}$.
Then we expect $g_{nm}$ to be constant for a fixed value of $g_4$.

In order to test the consistency of the FSS form for $f_s$, \eq{free}, we
measured the normalized moments $M_6=<P^6>/<P^2>^3$ and
$M_8=<P^8>/<P^2>^4$. From Eq. (A.12) and (A.13) we see that if $g_6$ and
$g_8$ are constant then $M_6$ and $M_8$ should also be constant. The
measured values of these moments together with the size of a typical error
are given in \table{gnm}. We see that $M_6$ and also $M_8$ are very stable
for the different lattice sizes, thus supporting our FSS ansatz for the
singular part of the free energy density.

\midinsert
$$
\vbox{\offinterlineskip
\halign{
\strut\vrule     \hfil $#$ \hfil  &
      \vrule # & \hfil $#$ \hfil  &
      \vrule # & \hfil $#$ \hfil  &
      \vrule # & \hfil $#$ \hfil
      \vrule \cr
\noalign{\hrule}
  ~~~N_\sigma~~~
&&~~~N_\tau~~~
&&~~~M_6~~~
&&~~~M_8~~~\cr
\noalign{\hrule}
 ~~8~&&~~4~&&~3.13~&&~6.94~\cr
 ~12~&&~~4~&&~3.11~&&~6.82~\cr
\noalign{\hrule}
 ~16~&&~~8~&&~3.16~&&~7.17~\cr
 ~24~&&~~8~&&~3.13~&&~6.97\cr
\noalign{\hrule}
 ~32~&&~16~&&~3.19(10)~&&~7.39(30)~\cr
\noalign{\hrule}}}
$$
\centerline {\bf \table{gnm}: \rm Normalized moments at $g_4=-1.4$.}
\bigskip
\bigskip
\endinsert
\noindent
{\bf Appendix B: The effective coupling scheme}
\medskip
In the $\beta_E$ scheme
the bare coupling, $\beta$, is replaced by an effective coupling,
$\beta_E$, which is related to the internal energy.
The starting point is a perturbative weak coupling
expansion of the internal energy, which is known up to
$O(\beta^{-3})$ \refs{heller}{pana},
\footnote{*}{We thank H. Panagopoulos
for informing us about the result of his $O(\beta^{-3})$ calculation
prior to publication.}
$$\eqalignno{
\langle U_p\rangle&=1-c_1\beta^{-1}-c_2\beta^{-2}-c_3\beta^{-3}
    +O(\beta^{-4})                                  & (\rm{B}.1) \cr
               c_1&=(N^2-1)(1/4)                    & ~~         \cr
               c_2&=(N^2-1)N^2(0.0204277-1/(32N^2)) & ~~         \cr
               c_3&=(N^2-1)N^4(4/3)
           (0.0066599-0.020411/N^2+0.0343399/N^4)~~.&            \cr}
$$
\noindent
An alternative coupling $\beta_E$ can now be defined from the first
two terms of $\langle U_p\rangle$,
$$ \eqalignno{
\beta_E &= {c_1\over 1-\langle U_p \rangle}~~,~~
   & (\rm{B}.2) \cr
             }
$$
with $U_p$ denoting the plaquette operator as defined in the third section.
The value of $\langle U_p\rangle$ has to be
determined by numerical simulations.
Our data for the average action are listed in \table{su2action}.
For $SU(3)$ the data of the average action
were taken from large symmetric lattices
\hbox{\hfil\refsss{stra}{has}{attig}{bali}.\hfil}
They are summarized in \table{su3action}.

\midinsert
$$
\vbox{\offinterlineskip
\halign{
\strut\vrule     \hfil $#$ \hfil  &
      \vrule # & \hfil $#$ \hfil
      \vrule \cr
\noalign{\hrule}
  ~\beta~
&&~1-\langle U_p \rangle ~\cr
\noalign{\hrule}
 ~~~~1.8800~~~~&&~~~~0.52637(5)~~~~\cr
 ~~~~2.1768~~~~&&~~~~0.43158(3)~~~~\cr
 ~~~~2.2986~~~~&&~~~~0.39746(1)~~~~\cr
 ~~~~2.3726~~~~&&~~~~0.37661(2)~~~~\cr
 ~~~~2.4265~~~~&&~~~~0.36352(1)~~~~\cr
 ~~~~2.5115~~~~&&~~~~0.34564(1)~~~~\cr
 ~~~~2.7395~~~~&&~~~~0.30869(2)~~~~\cr
\noalign{\hrule}}}
$$
\centerline {\bf \table{su2action}: \rm Expectation value of the average
                                        action for $SU(2)$.}
\bigskip
\endinsert
\midinsert
$$
\vbox{\offinterlineskip
\halign{
\strut\vrule     \hfil $#$ \hfil  &
      \vrule # & \hfil $#$ \hfil
      \vrule \cr
\noalign{\hrule}
  ~\beta~
&&~1-\langle U_p \rangle ~\cr
\noalign{\hrule}
 ~~~5.40~~~&&~~~0.52823( 30)~~~\cr
 ~~~5.51~~~&&~~~0.50120(100)~~~\cr
 ~~~5.60~~~&&~~~0.47520(~~2)~~~\cr
 ~~~5.70~~~&&~~~0.45100( 80)~~~\cr
 ~~~5.75~~~&&~~~0.44105(~~9)~~~\cr
 ~~~5.80~~~&&~~~0.43236(~~5)~~~\cr
 ~~~5.90~~~&&~~~0.41825(~~6)~~~\cr
 ~~~6.00~~~&&~~~0.40626(~~2)~~~\cr
 ~~~6.10~~~&&~~~0.39592(~~3)~~~\cr
 ~~~6.20~~~&&~~~0.38635(~~1)~~~\cr
 ~~~6.30~~~&&~~~0.37788(~~1)~~~\cr
 ~~~6.40~~~&&~~~0.36935(~~1)~~~\cr
 ~~~6.60~~~&&~~~0.35438(~~4)~~~\cr
 ~~~6.80~~~&&~~~0.34078(~~1)~~~\cr
 ~~~8.00~~~&&~~~0.27935(~~2)~~~\cr
\noalign{\hrule}}}
$$
\centerline {\bf \table{su3action}: \rm Expectation value of the
                                        average action for $SU(3)$.}
\bigskip
\endinsert
The normalization in Eq.~(B.2) is chosen such that
the leading term in the weak
coupling expansion of $\beta_E=2N/g_E^2$ is identical to the bare coupling,
$$ \eqalignno{
\beta_E &= \beta - c_2/c_1 + O(\beta^{-1}) ~~.~~ & (\rm{B}.3) \cr
            }
$$
Both couplings $g$ and $g_E$ agree in the limit when $g$ goes to zero
$$ \eqalignno{
g_E^2 &= g^2 +(c_2/c_1)g^4/(2N) +(c_3/c_1)g^6/(2N)^2 + O(g^8) ~~.~~
   & (\rm{B}.4) \cr
            }
$$
Additionally in both schemes the $\beta$-function
$a dg/da$ starts with the two universal coefficients
$b_0$ and $b_1$, which can be seen from an expansion
in the bare coupling $g$
$$ \eqalignno{
a{{dg_E}\over{da}}&=-b_0g^3  -b_1g^5+(3/2)(c_2/c_1)b_0g^5+O(g^7)&~\cr
                  &=-b_0g_E^3-b_1g_E^5 ~~.~~ & (\rm{B}.5) \cr
             }
$$
The constant $c_2/c_1$, however, redefines the lattice
$\Lambda$-parameter in the
$\beta_E$-scheme relative to the $\beta$-scheme,
$$ \eqalignno{
{\Lambda_E \over \Lambda_L} &= \exp{\bigl({{c_2/c_1}\over 4Nb_0}\bigr)}
                                 ~~.~~& (\rm{B}.6) \cr
             }
$$
This yields a much larger $\Lambda$-parameter for the alternative
coupling scheme,
$$ \eqalignno{
{\Lambda_E\over\Lambda_L} &= \cases{
1.7217 &, SU(2) \cr
2.0756 &, SU(3) \cr}
                      ~~.~~& (\rm{B}.7) \cr
 }
$$
\noindent
The $\beta$-function in the $\beta_E$ scheme can be separated into two
terms
$$ \eqalignno{
a{{dg_E}\over{da}}&=a{{dg}\over{da}} {{dg_E}\over{dg}}  & (\rm{B}.8) \cr
{{dg_E}\over{dg}} &={1\over{g_Eg^3}} {{16N^2}\over{N^2-1}}
                                     {{dE}\over{d\beta}}~~.~~& (\rm{B}.9) \cr
             }
$$

The first term is the usual $\beta$-function while the second term
$dE/d\beta$ is proportional to the ``specific heat'' $c_v$.
\footnote{*}{Here we consider the system as a $(d+1)$ dimensional model at
``inverse temperature'' $\beta$, which is not to be confused with the
physical temperature $T$.} The entire function $dg_E/dg$
asymptotically approaches unity  as the coupling $g$ goes to zero.

For the $SU(2)$ gauge theory a peak of the ``specific heat''
has been observed at $\beta \simeq 2.2$ \ref{laut} which can
be related to a nearby singularity in a generalized two
parameter coupling space \ref{BC}. For $SU(3)$
the same structure was found in the form of a huge bump in the
region $5.2 \le \beta \le 5.8$ \ref{columbiab}.

As a consequence the dip in the discrete
$\beta$-function of the bare coupling
seems to be compensated by the peak of the ``specific heat'',
which motivates the choice of $g_E$ as a better behaved coupling.

\vfil\end